\documentclass[11pt,a4paper]{article}
\pdfoutput=1

\usepackage{jheppub}

\usepackage{physics}
\usepackage{slashed}

\usepackage{microtype}

\usepackage{float}
\usepackage[export]{adjustbox}

\title{
A Sum Rule for Boundary Contributions to the Trace Anomaly
}
\author{Christopher P. Herzog, Vladimir Schaub}
\emailAdd{christopher.herzog@kcl.ac.uk}
\emailAdd{vladimir.schaub@kcl.ac.uk}
\affiliation{Mathematics Department, King's College London, \\
The Strand, London,  WC2R 2LS, UK }
\abstract{
In the context of boundary conformal field theory, we derive a sum rule that relates two and three point functions of the 
displacement operator.  For four dimensional conformal field theory with a three dimensional boundary, this sum rule
in turn relates the two boundary contributions to the anomaly in the trace of the stress tensor.
We check our sum rule for a variety of free theories and also for a weakly interacting theory, where a free scalar in the 
bulk couples marginally to a generalized free field on the boundary.  
}

\makeatletter
\def\@fpheader{\vspace{0cm}}
\makeatother

\begin{document}
\maketitle

\section{Introduction}

This work is about uncovering universal structure in conformal field theories (CFTs) with boundaries and defects.  
Such theories play an important role in diverse areas of theoretical physics.  The most straightforward application is to materials
that undergo phase transitions, for example magnetic ordering transitions close to the surface of  iron-aluminum alloys or nickel
\cite{Diehl:1997vr}.  There are however many other intriguing connections.  AdS/CFT correspondence in some of its simplest incarnations can be reformulated
after a Weyl transformation as a defect or boundary CFT \cite{Carmi:2018qzm,Herzog:2019bom}.  
Defect and boundary CFTs provide simple many-body examples where 
quantum entanglement can be efficiently computed \cite{Bianchi:2015liz}.  
Indeed, the entangling surface itself, through the replica trick, is often usefully reinterpreted as a codimension 
two defect.
The central result of this paper is a sum rule in boundary CFT that relates the two and three point functions of the displacement operator, an operator
that is always present in boundary CFT, dual to the location of the boundary.

We focus on the displacement operator because of its connection to anomalies in the trace of the stress tensor. That the breaking of scaling symmetry is anomalous means these trace contributions have the potential to provide non-perturbative information about the behavior of boundary and defect CFTs. Focusing on the particular example of a four dimensional CFT with a three dimensional boundary, Wess-Zumino consistency restricts the trace anomaly to have the schematic form  \cite{Herzog:2015ioa,Fursaev:2015uy,Solodukhin:2016tf} 
\begin{equation}
	\expval{T^{\mu}{}_{\mu}}=\frac{1}{16\pi^2}\left(c W_{\mu\nu\rho\lambda}^2-a E_{4}\right)+\frac{\delta(x_{\perp})}{16\pi^2}\left(a E^{(bry)}_{4}-b_1 [\hat{K}^3]-b_2 h^{\alpha\gamma}\hat{K}^{\beta\rho}W_{\alpha\beta\gamma\rho} \right) \ .
\end{equation}
Here $W_{\mu\nu\lambda\rho}$ is the Weyl curvature, $E_4$ is the Euler density along with its boundary contribution $E^{(bry)}_4$, while $K_{\mu\nu}$ is the extrinsic
curvature. By placing a hat on $K_{\mu\nu}$, we remove the trace, and $h_{\mu\nu}$ is a projector onto the boundary, located at $x_\perp = 0$. The constants $c$ and $a$ are the well known bulk anomaly coefficients.  The numbers $a$ and $c$ together specify the form of the stress-tensor two and three point functions \cite{Osborn:1994wz}. The number $a$ further has a number of special properties, useful in classifying CFTs.  It is monotonically decreasing under renormalization group flow \cite{Komargodski:2011vj} and independent of marginal couplings \cite{Osborn:1991gm}.  It is then worth asking what special properties, if any, the new boundary coefficients $b_1$ and $b_2$ may have.

It was recently shown \cite{Herzog:2017vj,Herzog:2018va} that $b_1$ and $b_2$ control correlation functions of the displacement operator. Consider the special case of a flat half space. The presence of a boundary affects the conservation of the stress tensor:
\begin{equation}
	\partial_{\mu}T^{\mu\nu}(x)=\delta(x_{\perp})n^{\nu}D(x) 
	\ ,
\end{equation}
where $n^\mu$ is an inward pointing unit normal vector along the boundary and $D(x)$ is the displacement operator. Integrating this expression on a pillbox that includes the boundary, one finds an equivalent representation of the displacement operator as the boundary limit $\lim_{x\cdot n \rightarrow 0} T^{nn}=D(x)$ of the normal-normal component of the stress tensor. The two and three point functions of the displacement operator are fixed by conformal symmetry up to constants, which are the boundary anomaly coefficients discussed above:
 \begin{equation}
 \label{b1b2def}
\begin{aligned}
		C_{\expval{DD}}&= \frac{15}{2\pi^4}b_2 \ , \; \; \;
		C_{\expval{DDD}}&=\frac{35}{2\pi^6}b_1 \ .
\end{aligned}
\end{equation}

In the case of $a$ and $c$,  the difference $a-c$ is bounded both above and below by a positivity constraint on the stress tensor \cite{Hofman:2008ar}. Theories that saturate the bound are free.  While the original argument invoked positivity of the energy in a conformal collider thought experiment, subsequent work rederived the bound in a more rigorous framework using causality \cite{Hartman:2016lgu} and reflection positivity plus crossing symmetry \cite{Hofman:2016awc}. A natural question is whether $b_1$ and $b_2$ may be bounded in a similar way.  Reflection positivity clearly demands $b_2>0$ \cite{Herzog:2017vj}, but could there be stronger bounds?

In this work, we take a step toward constraining these boundary anomaly coefficients.  We show that $b_1$ and $b_2$ satisfy a sum rule. Their difference is determined by a sum over boundary spin two operators that contribute to the boundary operator product expansion of the displacement-displacement-stress tensor three point function $\langle T_{\mu\nu}(x_1) D({\bf x}_2) D({\bf x}_3) \rangle$. This sum rule comes from a novel Ward identity that we derive. The identity gives us an equivalent way of expressing the sum rule, giving $c_{\expval{DD}}$ as an integral over the three point function $\langle T_{\mu\nu}(x_1) D({\bf x}_2) D({\bf x}_3) \rangle$.  While the connection to $b_1$ and $b_2$ is special to the 4d/3d set-up, our sum rule holds in general dimension.  

We are able to check that the sum rule is correct both for a selection of free theories and also for a weakly interacting  boundary CFT. The weakly interacting theory that we have in mind is free in the bulk with interactions confined to the boundary  \cite{Herzog:2017vj,Behan:2020vy}. In particular, we couple a free scalar in the bulk to a generalized free field on the boundary through a marginal interaction $g$.  A way of realizing such a situation is through a vector large $N$ limit.  One can introduce $N$ scalar fields $ \phi$ on the boundary that couple to the bulk field $\Phi$ via $g \phi^2 \partial_\perp \Phi $.  While the coupling $g$ will develop a beta function, the beta function vanishes in a large $N$ limit.  The generalized free field of the previous discussion is then $\sigma =  \phi^2$.  Replacing $ \phi$ with boundary fermions, $\psi$, the story can be repeated for a coupling $g \Phi \bar \psi \psi$.  These types of theories were discussed in detail recently in ref.\ \cite{Di-Pietro:2021wb}, including cases where the boundary starts out strongly interacting instead of weakly interacting and the generalized free field is the Hubbard-Stratonovich auxiliary scalar. There is, however, an even simpler realization of this type of interaction.  One can start with a pair of free fields in the bulk $\Phi_D$ and $\Phi_N$, one satisfying Neumann and one satisfying Dirichlet boundary conditions.  The marginal coupling is then $g \Phi_N \partial_{\perp} \Phi_D$, where either the boundary value of $\Phi_N$ or $\partial_{\perp} \Phi_D$ can be thought of as the generalized free field, depending on  one's point of view.

To perform the perturbative calculation and write our sum rule, we make extensive use of the embedding space formalism 
for CFT \cite{Costa:2011wa,Weinberg:2010ws,Liendo:2013tw}, adapted to our boundary context.\footnote{%
 A similar embedding space approach to conformal perturbation theory in the absence of a boundary can be found here \cite{Sen:2018tx}. For a general treatment of the embedding formalism applied to defects of generic codimension, see  \cite{Billo:2016vm,Lauria:2019wt,Guha:2018to}.
 }  
Embedding space works by lifting the theory into a space with two extra dimensions,
linearizing the action of the conformal group. We found that several of the integrals that we needed to do were
more easily done in this embedding space, adapting the approach of \cite{Simmons-Duffin:2014wb}.  
We hope some of the formalism we develop in sections \ref{sec:pert}, \ref{sec:block}, and appendix \ref{app:Int}
may be useful in related contexts.

The paper is structured as follows: In section \ref{sec:corr}, we recapitulate the construction of correlation functions of one bulk field and up to two boundary fields, using the embedding formalism. In section \ref{sec:ward}, we derive the Ward identity of CFT and boundary CFT, and comment on the constraint this gives on the CFT data. In section \ref{sec:pert}, we check our identity for a  free scalar theory and also for a non-trivial interacting example, where we set up conformal perturbation theory in embedding space. In section \ref{sec:block}, we discuss the conformal block expansion of  $\expval{TDD}$ and use it to rewrite the Ward identity as an algebraic constraint on the CFT data. Finally, we close with a discussion of our result.  Appendices contain details of how to carry out conformal integrals and also further examples of free theories,
where the Ward identity is explicitly checked.

\section{Classifying Correlators}\label{sec:corr}

The central characters of this work are two and three point correlation functions in boundary conformal field theory.  To set the stage, we review
how conformal symmetry constrains the form of these correlation functions using the framework of the embedding space method \cite{Weinberg:2010ws,Costa:2011wa}.  
More specifically, we consider bulk-boundary, bulk--bulk two point functions as well as bulk--boundary--boundary three point functions. We further only look at insertions of symmetric traceless tensors. The nomenclature bulk and boundary refers to where the operators are inserted, either on the boundary or in the bulk.
The bulk--boundary two point functions are fixed up to constants by conformal symmetry, while the remaining types will in general depend on functions of invariant cross ratios, constructed from the insertion locations. 

We include this section largely to set notation and conventions, but all its content can be inferred from \cite{Liendo:2013tw,Herzog:2017vj,Billo:2016vm}. Those familiar with the embedding space can skip this section, glancing at the main characters (\ref{FJGhatL}), (\ref{bulkbulk}), and (\ref{eq:corrT}) before moving to section 3.

\subsection{Embedding Space and Projection: \textit{A Lightcone Review}}

The embedding space is useful because it linearizes the action of the conformal group. We analyse a CFT living on $\mathbb{R}^{d}$ by lifting coordinates and operators to the projective light-cone of $\mathbb{R}^{d+1,1}$ \cite{Weinberg:2010ws,Costa:2011wa,Liendo:2013tw,Billo:2016vm}. The embedding space is defined as: 
\begin{equation}
	\frac{\{P^{A}\in \mathbb{R}^{d+1,1} \mid P^2 = 0\}}{ \{ P \sim \lambda P \mid \lambda \in \mathbb{R}^{+} \}} \ . 
\end{equation}
Sections of this space define a map to $\mathbb{R}^{d}$ on which the action of  Lorentz transformation pushes forward into a conformal mapping of $\mathbb{R}^{d}$. The generators of the conformal group in real space are given by the relevant push forward of the generators of Lorentz transformation on the embedding space:
\begin{equation}
	\mathcal{L}_{AB}=P_A\pdv{}{P_{B}}-P_B\pdv{}{P_{A}}+\mathcal{S}_{AB} \ ,
\end{equation}
with $\mathcal{S}_{AB}$ the spin part of the generator, that depends on the representation of the field it is acting on. We uplift operators $\mathcal{O}(x)$ to operators $\mathcal{O}(P)$, and then obtain real-space results by restricting to the corresponding section of the light-cone. For our purposes, we will focus on the Poincar\'e section, which maps to $\mathbb{R}^{d}$ in Cartesian coordinates, given by the parametrization: 
\begin{equation}
	(P^{+}, P^{-}, P^{\mu})=(1,x^2,x^{\mu}) \ .
\end{equation}

Conformal primary operators are identified by their conformal weight $\Delta$ and representation under the rotation group $SO(d)$.  
These operators in the embedding space obey the scaling relation: 
\begin{equation}
	\mathcal{O}( \lambda P )= \lambda^{-\Delta} \mathcal{O}(P ) \ .
\end{equation}
Tensor operators, $F_{\mu_1 \ldots \mu_J}(x)$ in real space can be thought of as the pull-back of tensor operators $F_{A_1 \ldots A_J}(P)$ in embedding space. The supplementary polarizations are set to zero by the pull-forward map, which allows us to look only at purely transverse tensors. In more detail, tensorial operators are uplifted through: 
\begin{equation}
\begin{aligned}
	F_{\mu_1 \ldots \mu_J}(x) &= \prod_{i=1}^{J}\pdv{P^{A_{i}}}{x^{\mu_i}}F_{A_1 \ldots A_J}(P(x)) \ , \\
	P^{A_i}F_{A_1 \ldots A_J}(\lambda P)&= 0  \ .
\end{aligned}
\end{equation}

Manipulation of tensorial operators can be simplified through the use of an index-free procedure \cite{Costa:2011wa}. This procedure can be performed analogously in real and embedding space. For symmetric traceless tensors (STT), we introduce a set of commuting auxiliary vectors $z^{\mu}$, respectively $Z^{A}$, and contract every free index of $F$ with $z$, respectively $Z$.\footnote{
	Operators which are not STTs can also be accommodated, by freeing the contracted indices of the tensor structure, and then 
	building all products of them with the right 	
	number of indices and with the symmetries specified by the Young Tableau of the operator, or using Grassmannian polarisation vectors \cite{Costa:2014rya,Guha:2018to,Lauria:2019wt}.} The result are scalar objects $F(x,z)$ and $F(P,Z)$. In real space, we impose $z^2=0$, while in embedding space we impose both $Z^2=0=Z\cdot P$. $F(P,Z)$ is then homogeneous of degree $J$ in $Z$, and so will be all of its correlation functions. Moreover, the transversality requirement is now formulated as $F(P,Z+\alpha P)=F(P,Z)$. The restriction of $F(P,Z)$ to the Poincar\'e section gives back $F(x,z)$ provided that $Z$ takes the form: 
\begin{equation}
	Z=(0,2x\cdot z, z^{\mu}) \ .
\end{equation}

Correlation functions in embedding space are constructed from the set of pairs  $(P_{m}, Z_{m})$, one for each insertion of an operator, subject to the constraints $P_{m}^2=Z_{m}^2=Z_{m}\cdot P_{m}=0$. Structures allowed by the conformal symmetry are Lorentz-invariant objects with appropriate scaling of the position vectors $P_m$ of each field, and appropriate homogeneity of the auxiliary parameters $Z_m$. An important building block in this procedure are the manifestly transverse structures $C_{AB}=P_{A}Z_{B}-Z_{A}P_{B}$, one for each insertion. 

Having constructed the correlation function in embedding space, 
the procedure of pulling back to real space is equivalent to a set of substitution rules for the contraction of these vectors:
\begin{equation}
	\begin{aligned}
		P_{m}\cdot P_{n}&=-\frac{1}{2}x_{mn}^{2} \  ,  \\
		P_{m}\cdot Z_{n}&= (x_m - x_n) \cdot z_n = x_{mn} \cdot z_n \ , \\
		Z_m \cdot Z_n &= z_m \cdot z_n  \ . 
	\end{aligned}
\end{equation}

Finally, we need to be able to retrieve tensor indices contracted with $z$ or $Z$. This can be done through the application of the Todorov differential operator: 
\begin{equation}
	D_{A}= \left(\frac{d-2}{2}+Z\cdot \frac{\partial}{\partial Z}\right)\pdv{}{Z^{A}} -\frac{Z_A}{2}\frac{\partial^2}{\partial Z \cdot \partial Z} \ . 
\end{equation}
In real space, the Todorov operator takes exactly the same form, with the same factors of $d$, with $Z$ replaced by $z$. After applying $D_{A}$, the result
can be rescaled to give  tensors with the usual normalisation. For example, freeing one index of a spin $J$, STT operator gives: 
\begin{equation}
	D_{A}F_{J}(Z,P)=\frac{Z^{A_2} \ldots Z^{A_{J}}F_{A_{1}\ldots A_{J}}(P)}{J\left(\frac{d}{2}+J-2\right)} \ ,
\end{equation}
where again, all factors of $d$ remain the same in embedding and real space, and the more general results can be found by induction. The conservation equation is expressed easily in this formalism by freeing one index and taking its divergence: 
\begin{equation}
	\partial^{\mu}D_{\mu} F = 0 \ .
\end{equation}
This Todorov operator also allows us to write the action of the spin-part of the Lorentz generators on our embedding space tensor fields:
\begin{equation}\label{eq:generator}
	\mathcal{L}_{AB}=P_A\pdv{}{P_{B}}-P_B\pdv{}{P_{A}}+\frac{Z_{A}D_B-Z_B D_A}{\frac{d}{2}+J-2} \ .
\end{equation}
The result follows from the fact that the tensor fields transform like a tensor product of $J$ times the vector representation.

We will be interested in theories with a boundary. The position of the boundary can be encoded through a vector $V^{A}$, which is simply a unit vector normal to  the boundary \cite{Liendo:2013tw}. The specificities of boundary CFT correlation functions arise from the existence of this additional vector to build correlation functions. Boundary operators naturally have $\hat{P}\cdot V=0$, while for bulk operators $P\cdot V \rightarrow x\cdot n$, $ Z\cdot V \rightarrow z\cdot n$. Similarly, we can define a restricted dot product, $P_1\bullet P_2 = P_1\cdot P_2 - (P_1\cdot V)(P_2\cdot V)$. Spinning boundary operators with position $P$ are accommodated by transverse polarisation vectors $W\cdot V=0=W^2=W\cdot P = P\cdot V$. We will sporadically write $p=d-1$.

With these preliminaries, we can use conformal symmetry to constrain boundary CFT correlation functions in the embedding space formalism, following \cite{Liendo:2013tw}. As previously stated, we restrict to the case of symmetric traceless tensor operators.

\subsection{Bulk--Boundary Two Point Function} 
\label{sec:bulkbrytwopoint}

We consider a generic two point function of a bulk spin $J$ operator $F_{J}(P_1,Z)$ and boundary spin $L$ operator $\hat{G}_{L}(P_2,W)$. Owing to the available building blocks, the possible tensor structures are given by: 
\begin{equation*}
	\begin{aligned}
		S_1 &= \frac{P_1\cdot V P_2\cdot Z}{P_1\cdot P_2}-Z\cdot V \ , \\
		S_{12}&= \frac{ P_1\cdot W P_2\cdot Z}{P_1\cdot P_2}-Z\cdot W \ .
	\end{aligned}
\end{equation*}
These two structures are related to the inversion bi-tensor, $I_{\mu\nu} = \delta_{\mu\nu} - \frac{2 x_\mu x_\nu}{x^2}$ \cite{McAvity:1995tm},  either contracted with the normal vector or projected in the tangent space to the boundary. The correlation function of such operators is fixed to be of the form
\begin{equation}
\label{FJGhatL}
	\expval{F_{J}(P_1,Z)\hat{G}_{L}(P_2,W)}=C_{\expval{F_J\hat{G}_{L}}}\frac{S_1^{J-L}S_{12}^{L}}{(P_1\cdot V)^{\Delta_1-\Delta_2}(-2P_1\cdot P_2)^{\Delta_2}} \ ,
\end{equation}
where the constant $C_{\expval{F_J\hat{G}_{L}}}$ is the only degree of freedom left unfixed by conformal symmetry.
To be able to convert this embedding space expression to a tensor in real space, we require $J\geq L$, consistent with the symmetry breaking pattern of $SO(d)\rightarrow SO(d-1)$. 
We also see the familiar restriction that only bulk scalar operators can have nonzero one point functions in boundary CFT.  
Indeed, for the boundary identity operator, with $L=0$ and $\Delta_2=0$, we find a one point function for the corresponding bulk operator.
This one point function must then be $P_2$ independent, which only happens when $J=0$.

For conserved bulk operators, conservation implies $\Delta_1 = d-2+J$ along with the restriction $\Delta_2=d$ for $J>L$, while $J=L$ is unconstrained \cite{Herzog:2017vj}. We further note however that for $\Delta_2 = d-2+L-1$, the boundary operator being shortened implies the vanishing of the two point function coefficient,
which will be important later in the discussion of the boundary stress tensor $\tau_{ij}$. Short boundary operators can however still couple to bulk operators in more general representations of the Lorentz group; for example, an electromagnetic field strength in the bulk can couple to a boundary conserved current \cite{Herzog:2017vj,DiPietro:2019hqe}.

The kinematic classification of these two point functions mirrors the restriction on the boundary operators appearing in the boundary operator product expansion (BOPE) of a bulk operator.

\subsection{Bulk--Bulk Two Point Function}

 The bulk--bulk two point function depends on the vectors $P_1$, $P_2$, $V$, $Z_1$ and $Z_2$. There exists a conformally invariant cross-ratio:
\begin{equation}
	u=\frac{(P_1 \cdot V)(P_2 \cdot V)}{-2P_1 \cdot P_2} \ .
\end{equation} 
For scalar operators, we can use conformal symmetry to fix the correlation function up to an unknown function of this cross-ratio:
\begin{equation}
	\expval{\mathcal{O}_1(P_1)\mathcal{O}_2(P_2)}=\frac{g(u)}{(P_1\cdot V)^{\Delta_1}(P_2\cdot V)^{\Delta_2}} \ .
\end{equation}

Adding spin to these operators is straightforward. The available objects are very similar to those that appeared for the corresponding bulk-boundary correlator.  
We have again $S_1$ and $S_{12}$, under the replacements $Z \to Z_1$ and $W \to Z_2$, as well as $S_2$, which is obtained from $S_1$ by exchanging the role of $P_1$, $Z_1$  and $P_2$, $Z_2$. For example, the correlator of a spin $J$ field and a scalar operator, both in the bulk, is fixed to be of the form
\begin{equation}
\label{bulkbulk}
	\expval{F_{J}(P_1;Z_1)\mathcal{O}(P_2)}=\frac{g(u)}{(P_1\cdot V)^{\Delta_1}(P_2\cdot V)^{\Delta_2}}\left(\frac{P_1\cdot V}{P_1 \cdot P_2}P_2 \cdot Z_1 - V \cdot Z_1 \right)^{J} \ .
\end{equation}
An important example is the $J=2$ case of a conserved stress tensor $T_{AB}$ \cite{McAvity:1993ul}. Indeed,
 for conserved operators, we can further apply the conservation constraint directly to obtain a differential equation for $g(u)$. For the $J=2$ case: 
\begin{equation}
\begin{aligned}
	(2 (d-2)u+d) g_T (u)-u (4 u+1) g_T'(u)&=0 \ .
\end{aligned}
\end{equation}
This equation is easily solved: 
\begin{equation}
	\begin{aligned}
		 g_{T}(u)&= C_{\expval{T\mathcal{O}}} \left(1+4u \right)^{\frac{d-2}{2}}\left(\frac{u}{1+4u} \right)^{d}  \ .
	\end{aligned}
\end{equation}
More generally, for conserved operators, the correlation function with a scalar is fixed up to a constant.

The bulk-bulk two point function can be expanded using conformal partial waves in two different ways -- by summing over intermediate bulk or boundary states. The equivalence of these two decompositions is a crossing symmetry constraint amenable to bootstrap analysis \cite{Liendo:2013tw}. 
This constraint for the scalar bulk-bulk two point function takes the form
\begin{equation}
	g(u)=\sum_{\hat{\Delta}} \frac{C_{\expval{\mathcal{O}_1 \hat{\mathcal{O}}}}C_{\expval{\hat{\mathcal{O}}\mathcal{O}_2}}}{C_{\expval{\hat{\mathcal{O}}\hat{\mathcal{O}}}}}\hat{g}_{\hat{\Delta}}(u)= \sum_{\Delta}C_{\expval{\mathcal{O}_1 \mathcal{O}_2 \mathcal{O}}}C_{\expval{\mathcal{O}}}g_{\Delta}(u) \ ,
\end{equation}
for known functions $\hat{g}_{\hat{\Delta}}(u)$, which are called boundary conformal blocks, and bulk conformal blocks $g_{\Delta}(u)$.

\subsection{Bulk--Boundary--Boundary Three Point Function}

In most of our computations, we will be interested  in a specific type of three point function  with one spinning bulk operator and two identical 
scalar boundary operators. In this configuration, we have two available tensor structures:

\begin{equation}
	\begin{aligned}
		Y_2 &= \frac{P_1\cdot V P_2\cdot Z}{P_1\cdot P_2}-Z\cdot V \ ,  \\
		Y_3 &= \frac{P_1\cdot V P_3\cdot Z}{P_1\cdot P_3}-Z\cdot V  \ .
	\end{aligned}
\end{equation}
The correlation function is not totally fixed by the residual conformal invariance, as there is a  cross-ratio: 
\begin{equation}
	v  = -\frac{1}{2}\frac{(P_1\cdot V)^2 P_2\cdot P_3}{(P_1\cdot P_2)(P_1\cdot P_3)} \ .
\end{equation}
The correlation function for a spin-$J$ bulk primary is now determined by summing over the possible polynomials in $Y_2 Y_3$ and $Y_2+Y_3$ with correct degree in $Z$, each multiplied by an unfixed function of the cross-ratio. For example, for a spin-$2$ operator we find: 
\begin{equation}\label{eq:corrT}
	\expval{F_2(P_1,Z)\hat{\mathcal{O}}(P_2)\hat{\mathcal{O}}(P_3)}=\frac{\alpha(v)(Y_2{}^2+Y_3{}^{2})+\beta(v)2Y_2 Y_3}{(-2P_1\cdot P_2)^{\frac{\Delta_1}{2}}(-2P_1\cdot P_3)^{\frac{\Delta_1}{2}}(-2P_2\cdot P_3)^{\Delta_2-\frac{\Delta_1}{2}}} \ .
\end{equation}
For the case  of the stress-tensor $T(P_1,Z)$ we have $\Delta_1=d$. Furthermore, 
 conservation implies a non-trivial differential constraint relating $\alpha(v)$ and $\beta(v)$: 
\begin{equation}
	0= d^2 \alpha_T (v)+\left(d^2-4\right) \beta_T (v)+2 (d (v-1)+1) \alpha_T '(v)+2 ((d-2) v-d+1) \beta_T '(v) \  .
\end{equation}

We will have the occasion to analyse this bulk--boundary--boundary three point function by bringing the bulk spin $J$ operator $F_J$ close to  the boundary and decomposing it as a sum over operators of spin $L \leq J$.  In this decomposition, we require the three point function of a boundary spin $J$ operator with two boundary scalars. It takes the well known form \cite{Costa:2011wa}
\begin{equation}
\label{hatGOO}
\begin{aligned}
	\expval{\hat{G}_L(P_1,Z)\hat{\mathcal{O}}(P_2)\hat{\mathcal{O}}(P_3)}&=C_{\expval{G_{L}\mathcal{O}\mathcal{O}}}\frac{V_{1;2,3}^{L}}{(-2P_1\cdot P_2)^{\frac{\Delta_1+L}{2}}(-2P_1\cdot P_3)^{\frac{\Delta_1+L}{2}}(-2P_2\cdot P_3)^{\Delta_2-\frac{\Delta_1-L}{2}}} \ , \\
	V_{1;2,3}&= -2(P_2\cdot P_1 Z \cdot P_3 - P_3 \cdot P_1 Z \cdot P_2) \ .
	\end{aligned}
\end{equation}
The factor of $2$ in the definition of $V_{i;j,k}$ is added to recover the canonical normalization of the three point function coefficient $C_{\expval{G_{L}\mathcal{O}\mathcal{O}}}$ in real space. We note that for identical scalar operators, Bose symmetry implies $L$ is even.

\section{Ward Identity}\label{sec:ward}

The goal of this section is to produce a relation between a correlation function of a string of operators that we write schematically
as $\langle \mathcal{X} \rangle$ and that correlation function with an extra insertion of the stress tensor $\langle T_{\mu\nu} (x) \mathcal{X} \rangle$.  
Such Ward identities are well known in the context of CFT but are less familiar once we add a boundary to the system \cite{McAvity:1993ul,Erdmenger:1997tr,Billo:2016vm}. 
We then apply this identity to the particular case where $\langle \mathcal{X} \rangle = \langle \hat {\mathcal O}(x_1) \hat {\mathcal O} (x_2) \rangle$
is a correlation function of two boundary scalars.  In fact, the case we are most interested in is where $\hat {\mathcal O}(x) = D(x)$ is the displacement
operator, because of its potential to shed light on boundary contributions to the anomaly in the trace
of the stress tensor. We begin by reviewing the usual Ward identity for CFT without a boundary and then add a boundary.

\subsection{Topological Operator in CFT}

Consider a coordinate transformation $x^{\mu} \rightarrow y^{\mu}$, and a concomitant Weyl transformation $g_{\mu\nu}\rightarrow  \Omega^{2}(x)g_{\mu\nu}$, such that the operator transforms as:
\begin{align}
	\mathcal{O}^{a}(x)&\rightarrow \mathcal{O}_i^{'a}(y_i)=\Omega(y_i(x_i))^{-\Delta_i}D\left[\pdv{y_i^{\mu}(x_i)}{x_i^{\rho}}
	\right]_{b}^{\quad a}\mathcal{O}_i^{b}(x_i(y_i)) \ .
\end{align}
We used $a$ and $b$ to denote abstract indices, and $D_{b}{}^{a}$ are representation matrices. Assuming that these are symmetries of the theory, from this transformation law, it follows that correlation functions transform in the opposite way: 
\begin{equation}\label{eq:trans}
\begin{aligned}
	\expval{\mathcal{O}_i^{a}(x_i)\ldots}_{g}&=\expval{\Omega(y_i(x_i))^{\Delta_i}D\left[\pdv{x_i^{\mu}(y_i)}{y_i^{\rho}}\right]_{b}^{\quad a}\mathcal{O}_i^{'b}(y_i(x_i))\ldots}_{\Omega^2 \tilde{g}} \ , \\
	\tilde{g}_{\mu\nu}&=g_{\rho\lambda}\pdv{x_i^{\rho}(y_i)}{y_i^{\mu}}\pdv{x_i^{\lambda}(y_i)}{y_i^{\nu}} \ .
\end{aligned}
\end{equation}

We consider a linearised transformation, under which we have:
\begin{equation}
\begin{aligned}
	\delta x^{\mu} = v^{\mu} \ , \; \;
	\Omega  =1+\omega \ , \; \; \delta g_{\mu\nu} = 2\omega  g_{\mu\nu}-\nabla_{\mu}v_{\nu}-\nabla_{\nu}v_{\mu} \ , \\
 \mathcal{O}^{'a}(y) -\mathcal{O}^{a}(x) \approx \delta \mathcal{O}^{a}(x) = -\Delta \omega(x)\mathcal{O}^{a}(x)-\mathcal{L}_{v} \mathcal{O}^{a}(x)  \ , \\
\end{aligned}
\end{equation}
where $\mathcal{L}_{v}$ is the Lie derivative. Now, let $\mathcal{X}$ to be a string of operator insertions. Comparing both sides of \eqref{eq:trans} and expanding for a linearized transformation, we relate the transformation of $\mathcal{X}$ to a change in the metric:
\begin{equation}
\begin{aligned}
\label{eq:wardish}
	\expval{\delta \mathcal{X}}_{g}&=\Big(\expval{ \mathcal{X}}_{g+\delta g}-\expval{ \mathcal{X}}_{g} \Big) \\
	&=\frac{1}{2}\int dV \expval{T^{\mu\nu} \mathcal{X}}_{g}\delta g_{\mu\nu} \\
	&=\int dV  (\omega g_{\mu\nu}+v_{\mu}\partial_\nu)\expval{T^{\mu\nu} \mathcal{X}}_{g} \ , 
\end{aligned}
\end{equation}
where we defined insertions of the stress tensor $T^{\mu\nu}$ as the response of the correlation function to a metric deformation. Note that taking $v^{\mu}$ to be a conformal Killing vector field, and $\omega(x)$ its associated weight, the correlation function is invariant, since $\delta g_{\mu\nu}=0$. This is simply the statement of conformal invariance. 

\begin{figure}[H]
	\centering
	\includegraphics[width=0.6\linewidth]{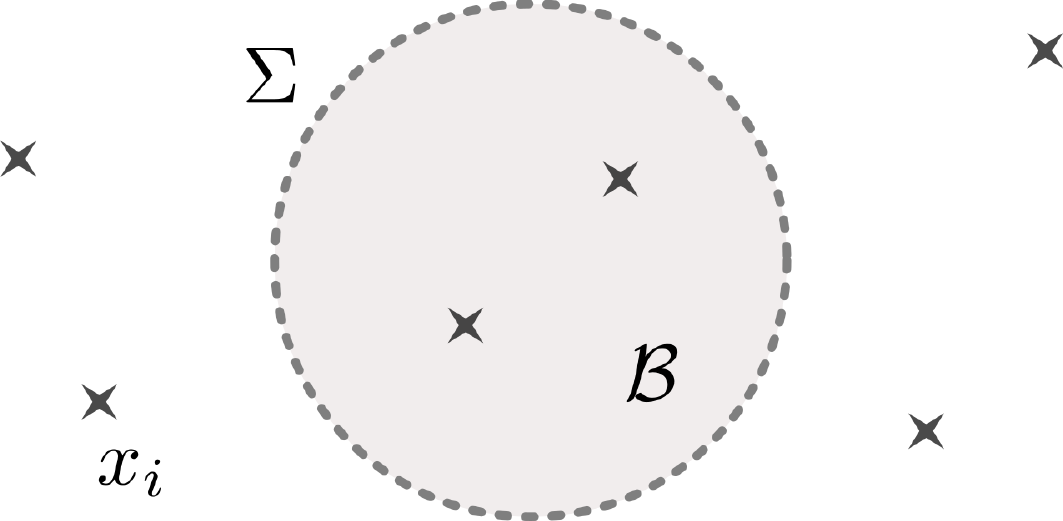}
    \caption{Typical situation we consider. There are numerous field insertions, with a shaded domain in which fields are transformed, bounded by the dashed region of insertion of $\mathcal{I}_\Sigma$.
    \label{fig:nobry}}
\end{figure}

By taking functional derivatives with respect to $\omega$ or $v^\mu$, we recover the usual contact-term form of the Ward identities. This form shows that they hold locally, and can be written as a transformation acting only on some subset of the fields. We prefer an integrated form for what follows and define a
topological operator to carry out the symmetry transformation inside a compact domain $\mathcal{B}$,
with boundary $\Sigma=\partial\mathcal{B}$.  The domain $\mathcal{B}$, which we can imagine to be a sphere or any topologically equivalent object,  surrounds some of the local operator insertions of ${\mathcal X}$, but does not contain any on its boundary. The situation is visualized in fig.~\ref{fig:nobry}.

We construct an operator living on this surface $\Sigma$, using the stress-tensor and a conformal Killing  vector field $v_{\mu}$:
\begin{equation}
	\begin{aligned}
		\mathcal{I}_{\Sigma}\left[\expval{\mathcal{X}}_g, v^{\mu}\right] &= \int_{\Sigma}d S^\mu v^\nu\expval{T_{\mu\nu} \mathcal{X}}_g =\int_{\mathcal{B}}dV \nabla^\mu\expval{v^\nu T_{\mu\nu}\mathcal{X}}_g \\
		&= \int_{\mathcal{B}}dV \expval{\left(\omega g^{\mu\nu} T_{\mu\nu}+v^\nu\nabla^{\mu} T_{\mu\nu}\right)\mathcal{X}}_g \\
		&=\expval{\left.\delta\right\rvert_{\mathcal{B}} \mathcal{X}}_{g}	\end{aligned}
\end{equation}
In the last line we denoted $\left.\delta\right\rvert_{\mathcal{B}}$ the variation $\delta$ acting only inside the domain $\mathcal{B}$. This operator is invariant under deformation of $\mathcal{B}$, as long as we do not cross an insertion, as is clear from its explicit action. It is indeed topological, and implements the symmetry transformations locally. 

In practice, this procedure brings us to an integrated form of the usual Ward identities. It is however a convenient form, since the smearing alleviates subtleties  associated with the matching of contact-terms. Two classic examples are the following. First, consider $\mathcal{B}$ to be the half space, separating operator insertions as $\mathcal{X}= \mathcal{X}_1 \mathcal{X}_2$, with $\mathcal{X}_1$ in $\mathcal{B}$, the others outside. This operator becomes an insertion of the conformal generator associated to $v^{\mu}$ in the separation: e.g.\   taking 
$v^{\mu}=a^{\mu}$ constant, $\mathcal{I}_{\Sigma}[\expval{\mathcal{X}_1 \mathcal{X}_2},a^{\mu}]=a^{\mu}\expval{\mathcal{X}_1 P_\mu \mathcal{X}_2}$. 
Second,
an important example \cite{Cardy:1987dg} is where $\langle \mathcal{X} \rangle = \expval{\phi(x)\phi(x')}$ is the correlation function of two scalar operators. Let $v^{\mu}\propto x^{\mu}$  be a dilation and $\Sigma$ a ball of unit radius. Applying the definition of $\mathcal{I}_{\Sigma}[\expval{\phi(0)\phi(\infty)},\lambda x^{\mu}]$, one finds a relation between the three point function coefficient $C_{\expval{T\phi\phi}}$ and the two point function coefficient $C_{\expval{\phi\phi}}$:
\begin{equation}
\label{TOOtoOO}
	\mathcal{I}_{\Sigma}[\expval{\phi(0)\phi(\infty)},\lambda x^{\mu}] \Rightarrow C_{\expval{T\phi\phi}}=-\frac{d\Delta}{(d-1)S_{d}}C_{\expval{\phi\phi}}
\end{equation}
where we defined $S_{d}\equiv \frac{2\pi^{\frac{d}{2}}}{\Gamma(\frac{d}{2})}$. Our goal is now to extend this discussion to a situation with a boundary. 

\subsection{Adding a Boundary}

We adapt our previous discussion to the presence of a boundary, offering a generalisation of the treatment of Ward Identity with a boundary of \cite{McAvity:1993ul}. Our starting point is the same as eq.\ \eqref{eq:wardish}: knowing the transformation property of correlation functions, we can trade deformation of the insertions for deformation of geometric quantities. We will consider a small deformation away from flat space and a planar boundary localised at $x\cdot n =0$. In this setup, the only geometric quantities necessary to describe the system are the metric and the embedding of the boundary in a given coordinate system. Hence, we  consider the response of the correlator to a metric variation as well as a general shift $\delta y = n^{\mu}\delta x_{\mu}$ in the position of the boundary:
\begin{equation}\label{eq:FirstVar}
	\expval{\delta\mathcal{X}}= \frac{1}{2}\int_{\mathcal{M}}dV\delta g_{\mu\nu}\expval{T^{\mu\nu}\mathcal{X}}+ 
        \int_{\partial\mathcal{M}}dS
	\left[ \frac{1}{2}\delta \gamma_{ij}\expval{\tau^{ij}\mathcal{X}}+\delta y \expval{D\mathcal{X}} \right] \ . \nonumber
\end{equation}
Here $\gamma_{ij}$ is the boundary metric. The operator $\tau_{ij}$ can be seen as transverse boundary localised contribution to $T_{\mu\nu}$; we however choose to make it explicit, as it turns out to be useful when discussing BCFT where the bulk and boundary are weakly coupled. The operator $D$ is the displacement operator, encoding the response of the correlator under a shift of the boundary. We are interested in the effect of conformal transformations close to flat space, which preserve a flat boundary. This setup is similar to the starting point of the derivations \cite{Billo:2016vm}. The difference is that they consider the response of the correlation function with respect to generic transformations, some of which are broken symmetries, while we are interested in constructing explicit topological operators implementing unbroken symmetries out of local insertions. These preserved conformal transformations leave the extrinsic geometry of the boundary invariant, and will naturally correspond to transformations for which $\delta y =0$. Analogously to what we obtained previously we find
\begin{equation}
\begin{aligned}
	\delta g_{\mu\nu}&= 2\omega \eta_{\mu\nu}-\partial_{\mu}v_{\nu}-\partial_{\nu}v_{\mu} \ , \\
	\delta \gamma_{ij}&=2\omega\eta_{ij}-\partial_{i}v_{j}-\partial_{j}v_{i} \ . \\
	\end{aligned}
\end{equation}
Meanwhile, the explicit transformation of $\mathcal{X}$ is computed as before. Using these relations, we can formulate the analogue of eq.\ \eqref{eq:wardish}: 
\begin{equation*}
\begin{aligned}
	\expval{\delta \mathcal{X}}&= \int_{\mathcal{M}}dV (\omega\eta_{\mu\nu}+v_{\mu}\partial_{\nu})\expval{T^{\mu\nu}\mathcal{X}}+\\
	& \hspace{1in} \int_{\partial\mathcal{M}}dS \biggl[ (\omega\eta_{ij}+v_{i}\partial_j)\expval{\tau^{ij}\mathcal{X}}-v_{\mu}\expval{T^{\mu n}\mathcal{X}}+v_ n \expval{D\mathcal{X}} \biggr] \ .  \end{aligned} 
\end{equation*}
From rotational symmetry, we have a splitting on the boundary $\left.v^{\mu}n^{\nu}T_{\nu\mu}\right \rvert_{\partial\mathcal{M}}=v^{n}T^{nn}+ v^{i}V_{i}$ which defines the vector flux $V_i$. Requiring that a full translation of the boundary, as well as all insertions of operators, along the broken direction leave the correlation function invariant, imposes $\left.T_{nn}\right \rvert_{\partial\mathcal{M}}=D$. Both $D$ and $V^{i}$ have protected dimension $d$, and they are naturally mapped to the allowed spin-$0$ and $1$ primaries in the boundary OPE of $T$. We see from our derivation that outside of insertion points, there is another operator equation, now relating $\tau$ to $V_i$: 
\begin{equation*}
	\partial_{i}\tau^{ij}= V^{j} \ .
\end{equation*}
This equation induces a recombination: without the bulk, $\tau_{ij}$ is a \textit{bona fide} stress tensor. However, the bulk theory furnishes a momentum flux operator which recombines with the short conformal family of $\tau$, lifting it to unprotected dimension and out of this discussion \cite{Herzog:2017vj,Behan:2020vy}. From this reasoning, we can gather 3 interesting scenarios: 

\begin{enumerate}
	\item Both $T$ and $\tau$ are present: This describes a non-recombined bulk-boundary theory. The theory has different orthogonal sectors that do not speak to each other. In the Ward identity we will derive, both $T$ and $\tau$ contributions are necessary in general. 
	\item Only $T$ is present: This is the generic boundary CFT setup we imagine, a recombined system. There is one sector englobing the bulk and boundary degrees of freedom, and the splitting between bulk and boundary response to metric variation was not meaningful.
	\item $T$ is absent, $\tau$ is present: This is a holographic setup. The bulk correlation functions are gravitational, while the boundary theory has a conserved stress tensor: it is local. 
\end{enumerate}

The first situation will be encountered in our perturbative computation, where coupling the bulk and boundary theories leads to
$\tau$ acquiring an anomalous dimension $\gamma \sim \mathcal{O}(g^2)$
Nevertheless, 
the Ward identity associated to ${\mathcal X} = D(x) D(x')$ is not sensitive to the anomalous dimension because the 3pt.\ function coefficient $C_{\expval{\tau DD}}$
is already $O(g^2)$. The second situation will be the focus of our generic investigation of the anomaly coefficients of boundary CFTs. The last one connects to the  AdS/CFT correspondence, and we leave it for future investigation.  The non-conservation of the boundary stress tensor is holographically
dual to the graviton getting a mass in anti-de Sitter space \cite{Aharony:2003qf,Aharony:2006hz}. 

Following this construction, we can now construct topological operators implementing locally the symmetries of the theory. We picture a generic domain $\mathcal{B}\in \mathcal{M}$ which englobes some operator insertions, both potentially in the bulk or on the boundary. The boundary of $\mathcal{B}$ splits naturally in two pieces: $\partial \mathcal{B}=\Sigma\cup \Sigma_{\partial}$, where $\Sigma_{\partial}=\partial \mathcal{B}\cap \partial \mathcal{M}$. Naturally, $\partial \Sigma  = \partial \Sigma_{\partial} = \Gamma$ is a codimension-1 submanifold of $\partial \mathcal{M}$. This is visualised as follows: 

\begin{figure}[H]
	\centering
  	\includegraphics[width=0.8\linewidth]{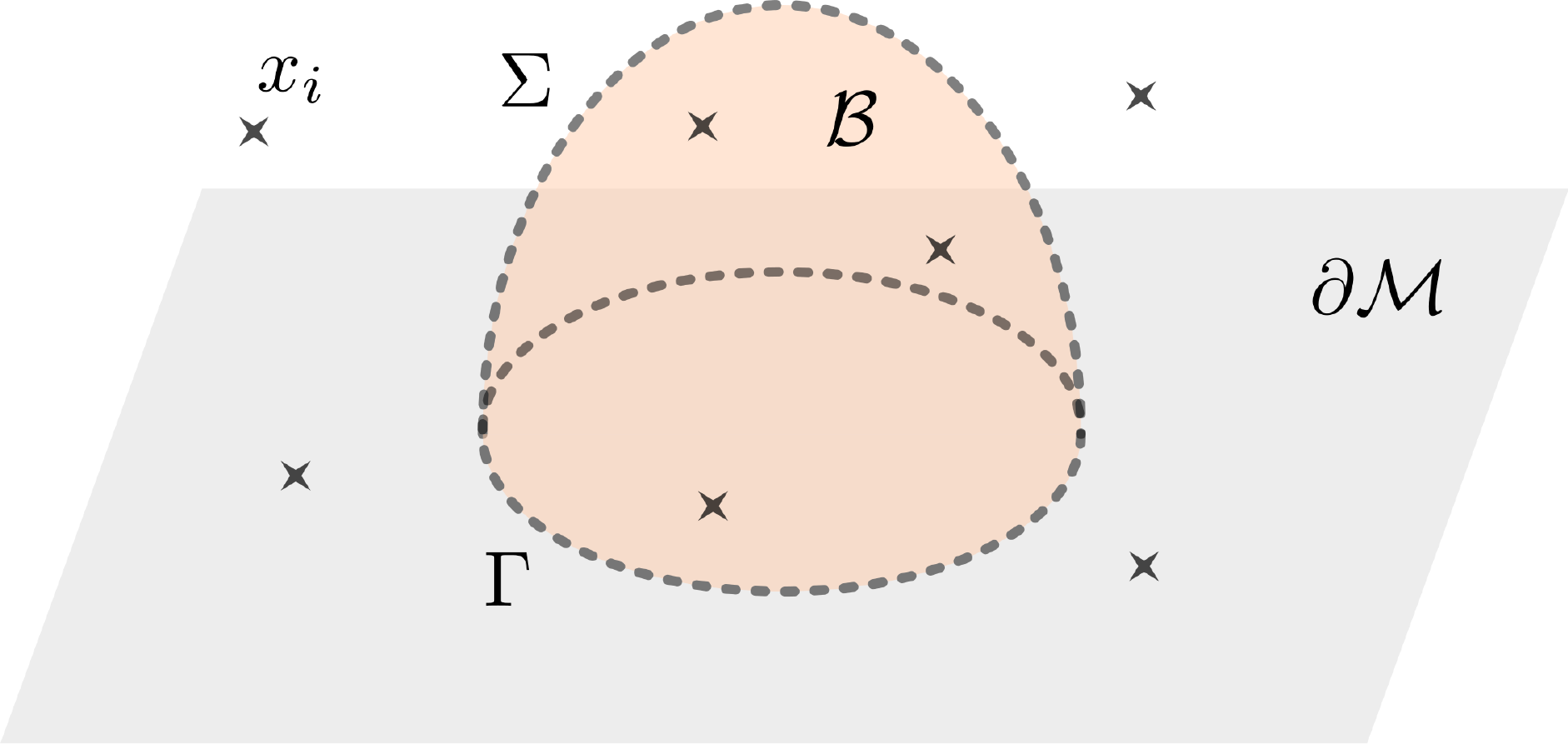}
    \caption{Typical situation with a boundary. The boundary is represented by the opaque plane. Field insertions in the bulk and boundary are given by stars. 
    The coloured domain is $\mathcal{B}$. The dashed lines correspond to the boundary of $\mathcal{B}$. 
    In the bulk, this is the surface $\Sigma$, 
    and on the boundary $\Gamma=\partial \Sigma$. $\mathcal{I}_{\Sigma}$ is inserted all along the boundary of ${\mathcal B}$.}
\end{figure}

We can now exhibit the topological operator $\mathcal{I}_{\Sigma}$ implementing conformal transformations given by the conformal Killing vector field $v$: 
\begin{equation}
\begin{aligned}
	\mathcal{I}_{\Sigma}\left[\expval{\mathcal{X}},v^{\mu}\right]&=\int_{\Sigma}  dS^{\mu}v^{\nu}\expval{T_{\mu\nu}\mathcal{X}}+\int_{\Gamma}  d\sigma^{i}v^{j}\expval{\tau_{ij}\mathcal{X}} \\
	&=\int_{\mathcal{B}} dV\partial^{\mu}(v^{\nu}\expval{T_{\mu\nu}\mathcal{X}})-\int_{\Sigma_{\partial}}dSv^{i}\expval{T_{ni}\mathcal{X}}+\int_{\Gamma}  d\sigma^{i}v^{j}\expval{\tau_{ij}\mathcal{X}} \\
	&=\int_{\mathcal{B}}dV(\omega g^{\mu\nu}+v^{\nu}\partial^{\mu})\expval{T_{\mu\nu}\mathcal{X}}+\int_{\Sigma_{\partial}}dS(\omega \gamma^{ij}+v^{i}\partial^{j})\expval{\tau_{ij}\mathcal{X}}-v^{\mu}\expval{T_{n\mu}\mathcal{X}} \\
	&=\expval{\left.\delta\right\rvert_{\mathcal{B}} \mathcal{X}}\ .
\end{aligned}
\end{equation}
This computation shows that this operator correctly implements conformal transformations locally, both  for bulk and boundary insertions, provided we consider only allowed transformations $\left. v_{n}\right\lvert_{\partial M}=0$. As an application of this procedure, we can consider $\expval{\mathcal{X}}=\expval{\mathcal{O}(x)}$, some generic bulk scalar operator, and $\mathcal{B}$ to be a ball of radius $\epsilon\ll 1$ around it. In this configuration, $\Sigma_{\partial}\equiv  \emptyset$ and we can compute directly both sides of the operator for a dilation $v^\mu = \lambda x^\mu$ to find the result 
\begin{equation*}
	C_{\expval{T\mathcal{O}}}=- 2^d \frac{d \Delta}{(d-1)} \frac{1}{S_d} C_{\expval{\mathcal{O}}} \ ,
\end{equation*}
where $C_{\expval{\mathcal{O}}}$ is the one point function coefficient of the operator $\mathcal{O}$. This result was obtained originally by other means in \cite{Cardy:1990xm,McAvity:1995tm}.

A more interesting application of the Ward identity is to the two point function of two identical boundary operators, $\langle \hat{\mathcal{O}}(x) \hat{\mathcal{O}}(x') \rangle$:
\begin{equation*}
	\mathcal{I}_{\Sigma}\left[\langle \hat{\mathcal{O}}(0) \hat{\mathcal{O}}(\infty) \rangle,\lambda x^{\mu}\right]=-\lambda \Delta \langle \hat{\mathcal{O}}(0) \hat{\mathcal{O}}(\infty) \rangle \ ,
\end{equation*}
with  $\Sigma$ a half-unit-sphere centred around the origin. The Ward identity translates into an integral constraint:
\begin{equation}\label{eq:ward}
\begin{aligned}
\frac{2}{d}S_{d-1}\int_{0}^{1}\Big[\alpha(x^2)\left(1-d x^2 \right)+\beta(x^2) &\left(1-x^2(2-d)\right)\Big]\sqrt{1-x^2}^{d-3}dx \\
&=\frac{d-2}{d-1}S_{d-1}C_{\expval{\tau \hat{\mathcal{O}} \hat{\mathcal{O}}}}+\Delta C_{\expval{\hat{\mathcal{O}} \hat{\mathcal{O}}}} \ .
\end{aligned}
\end{equation}
The functions $\alpha$ and $\beta$ are the ones appearing in the correlation function \eqref{eq:corrT}. 
Note that when the bulk $T_{\mu\nu}$ vanishes, the Ward identity reduces to the usual CFT result (\ref{TOOtoOO}).

In this work, we are most interested in the special case where these boundary operators are displacement operators. The functions $\alpha$ and $\beta$  can be expanded as a sum of conformal blocks $\alpha_{l,\Delta}(u)$ and $\beta_{l,\Delta}(u)$. In section \ref{sec:block} we will compute these functions, and rewrite this identity  as a constraint on the CFT data of $D$.
Optimistically, one might hope that such an integral relation might give a simple and direct relation between the
displacement correlation function coefficients, $C_{\expval{DDD}} \sim C_{\expval{DD}}$ (and hence between the corresponding
anomaly coefficients in the trace of the stress tensor).   Our analysis will show that generically to establish such a relationship
one needs to know the tower of spin two
operators that appear in the boundary OPE of the stress tensor, in addition to $C_{\expval{DD}}$.

\section{Conformal Perturbation Theory with a Boundary}\label{sec:pert}

In this section we provide explicit checks of the Ward identity \eqref{eq:ward} by recovering $C_{\expval{DD}}$ from $\langle T_{\mu\nu} D D \rangle$ for free 
and weakly interacting theories. In particular, we consider a free scalar in general dimension.  
(Appendix \ref{sec:free4d} contains results for  a free fermion and 4d Maxwell field.) 
We then add weak interactions to the scalar through a classically marginal perturbation on the boundary.
Introduced in ref.\ \cite{Herzog:2017vj} as examples of boundary CFT accessible to perturbation theory,
this class of theories was recently analyzed in great detail from a bootstrap approach \cite{Behan:2020vy} and in the large $N$ limit 
\cite{Di-Pietro:2021wb}. We take a covariant approach, reformulating the conformal perturbation theory in embedding space \cite{Simmons-Duffin:2014wb,Sen:2018tx,Fukuda:2018vi}.

\subsection{Free Scalar Field}

We look at the simplest non-trivial example on which to test our Ward identity, a single free scalar field in the presence of a boundary. The bulk equation of motion $\Box \phi = 0$ allows only two 
possible non-zero bulk-boundary two point functions $\langle \phi(x) \varphi_{1,2}({\bf x}') \rangle$  \cite{Lauria:2021ut} where
\begin{itemize}
	\item \(\varphi_1 \equiv \lim_{n\cdot x\rightarrow 0}\phi(x)\), with protected dimension \(\Delta_1 = \frac{d}{2}-1\) \ , 
	\item \(\varphi_2 \equiv \lim_{n\cdot x\rightarrow 0}\partial_n\phi(x)\), with protected dimension \(\Delta_2 = \frac{d}{2}\) \ .
\end{itemize}
Said another way, $\varphi_{1,2}$ are the only two possible operators in the boundary OPE of the bulk scalar $\phi$.
Clearly in the case of Dirichlet boundary conditions, $\varphi_1$ will be absent, while for Neumann, $\varphi_2$ is set to zero.

The bulk equations of motion also severely restrict the form of the bulk--bulk two point function:
\begin{equation}
\begin{aligned}
\label{phitwopt}
	\expval{\phi(x,y)\phi(0,y')} &=\kappa\left(\frac{1}{(x^2+(y-y')^2)^{\Delta_1}}+\frac{\chi}{(x^2+(y+y')^2)^{\Delta_1}}\right)  \ .
\end{aligned}
\end{equation} 
We have used the usual field theory normalization $\kappa^{-1} = (d-2)S_{d}$.
The reflection coefficient $\chi$ determines the type of boundary condition; $\chi = 1$ is Neumann and $\chi = -1$ is Dirichlet. In fact, we will be able to obtain more general values of $\chi$ in the next section through the addition of a boundary interaction.

In the bulk, we have a conserved stress-tensor given in index-free notation by: 
\begin{equation}
	T(x,z)=[(z\cdot \partial \phi(x))^2]+\frac{(d-2)}{4(d-1)}(z\cdot\partial)^2[\phi(x)^2] \ .
\end{equation}
By the square bracket we denote the usual point splitting procedure to remove the divergence in the coincident limit. From the  boundary limit of $T$, we find the displacement operator: 
\begin{equation}
	\lim_{x\cdot n\rightarrow 0}T^{nn}(x)\equiv D(x)=\frac{1}{2}[\varphi_2^2]-\frac{1}{2}[(\partial_i \varphi_1)^2]+\frac{d-2}{4(d-1)}\partial_i\partial^{i}[\varphi_1^2] \ .
\end{equation}

With these ingredients in hand, it is a straightforward matter to compute the correlation functions $\langle D({\bf x}) D({\bf x}') \rangle$
 and $\langle T_{\mu\nu}(x_1) D({\bf x}_2) D({\bf x}_3) \rangle$ and test the Ward identity.  
We are working with a free theory, and these correlation functions follow from Wick's Theorem and the
$\langle \phi \phi \rangle$ two point function (\ref{phitwopt}).  
For the displacement two point function, we find from direct calculation that 
\begin{equation}
\label{cDDactual}
C_{\expval{DD}} = (d-2)^2 \kappa^2 (1 + \chi^2)  \ .
\end{equation}
As we reviewed above, the expression (\ref{eq:corrT}) for $\langle T_{\mu\nu}(x_1) D({\bf x}_2) D({\bf x}_3) \rangle$ depends on functions $\alpha(v)$ and $\beta(v)$, which
here can be deduced from Wick's Theorem:
\begin{equation*}
	\begin{aligned}
		\alpha(v)&= -\frac{(d-2)^3 d \kappa ^3}{32 (d-1)^3 v}\Big(d^5 v^2 (\chi +1)^3+2 d^4 v \left(v (\chi +1)^3-3 \chi  ((\chi -1) \chi +3)+1\right) \\
		&+ d^3 \left(-12 v^2 (\chi +1)^3+4 v (\chi  ((\chi -9) \chi +3)-3)+3 (\chi +1)^3\right) \\
		&-4 d^2 \left(2 v^2 (\chi +1)^3 - 6 v \chi  \left(\chi ^2+3\right)+(\chi +1)^3\right) \\
		&+4 d \left(8 v^2 (\chi +1)^3-4 v (\chi  ((\chi -9) \chi +3)-3)+(\chi +1)^3\right)-32 \left(3 v \chi ^2+v\right)\Big) \ , \\
		\beta(v)&= \frac{(d-2)^3 d^2 \kappa ^3}{32 (d-1)^3 v}\Big(d^4 v^2 (\chi +1)^3+2 d^3 v \left(v (\chi +1)^3-3 \chi  ((\chi -1) \chi +3)+1\right) \\
		&+d^2 \left(-4 v^2 (\chi +1)^3+4 v (\chi  ((\chi -9) \chi +3)-3)+3 (\chi +1)^3\right)\\
		&-4 d \left(2 v^2 (\chi +1)^3-2 v (\chi  (\chi  (2 \chi +9)+6)+3)+(\chi +1)^3\right)-4 (4 v-1) (\chi +1)^3\Big) \ .
	\end{aligned}
\end{equation*}
That these expressions are polynomials of the form $av + b + c v^{-1}$ appears to be a standard feature of free theories.
The Maxwell field and massless fermion we consider in appendix \ref{sec:free4d} share this feature.

If we now insert this result for $\alpha$ and $\beta$ into the left hand side of the Ward identity (\ref{eq:ward}) and integrate, we find that
\begin{align}
\label{cDDpredicted}
\frac{2}{d^2}S_{d-1}\int_{0}^{1}  \left[\alpha(x^2)\left(1-d x^2 \right)+\beta(x^2)\left(1-x^2(2-d) \right)\right]&  \sqrt{1-x^2}^{d-3}dx   \\
&=
(d-2)^2 \kappa^2 \frac{(1 + 3 \chi^2)}{2}
 \ , \nonumber
\end{align}
which reproduces $C_{\expval{DD}}$ (\ref{cDDactual}) 
precisely in the case of Neumann $\chi = 1$ and Dirichlet $\chi=-1$ boundary conditions.
We conclude in these cases that $C_{\expval{\tau DD}}$ must vanish.
In the next section we will see how to recover the Ward identity for more general values of $\chi$ by introducing a classically marginal boundary
term that allows us to tune perturbatively away from exactly Dirichlet or Neumann boundary conditions.

\subsection{Perturbation Around the Free Boson}

We investigate the classically marginal deformation localized  on the boundary: 
\begin{equation}
	\expval{\mathcal{X}}_{g}=\expval{\mathcal{X}e^{g\int_{\partial \mathcal{M}}dS \, \varphi_i \sigma}} \ .
\end{equation}
The exponential is to be understood through its Taylor series. We often drop the $i$ and write expressions in  a form valid for both $1$ and $2$. Note $\sigma$ is an operator of dimension $p-\Delta_{\varphi}$, which under this perturbation joins the spectrum of boundary operators in the CFT.
The deformation causes the short multiplet of the boundary stress tensor $\tau$ to recombine with a vector operator (the boundary limit of the 
divergence of the bulk stress tensor) into a longer multiplet \cite{Herzog:2017vj}. 
In simpler terms, $\tau$ is no longer a conserved operator and develops an anomalous dimension.  
This recombination however will only affect our calculation at subleading orders in perturbation theory, and we will not see it.

Our goal is to check our Ward identity at leading order. 
At leading order, the main effect of the perturbation is to shift the value of $\chi$ in (\ref{phitwopt}) slightly away from Dirichlet $\chi = -1$ or Neumann $\chi = 1$. 
As a result, the effect on  $\expval{TDD}$ and $\expval{DD}$ can be captured through Wick contractions, using a general value of $\chi$.
Indeed, we already computed $\expval{TDD}$ and $\expval{ DD}$ for general $\chi$ in the previous section.
The remaining tasks are then twofold.  We must determine precisely how the perturbation shifts $\chi$ in $\expval{\phi\phi}$, and we must
determine $\expval{\tau DD}$, which will be nonzero in this perturbative framework.

We compute the effect of the coupling $g$ on $\chi$ and $\expval{\tau DD}$ using conformal perturbation theory and an embedding space approach
\cite{Simmons-Duffin:2014wb}.
The first task is then to uplift the perturbation to embedding space:
\begin{equation}
	\int_{\partial \mathcal{M}}d^{p}x \, \varphi(x) \sigma(x)\rightarrow  \int D^{p}X \, \varphi (X) \sigma(X) \ .
\end{equation}
The mechanics of how to compute with these integrals is reviewed in appendix \ref{app:Int}, 
where we also give the technical details of some of the calculations to follow. 

\subsubsection{Computation of $\expval{\phi\phi}$}

It is useful to lift $\expval{\phi\phi}$ to embedding space before we begin, and also to 
understand how the boundary operators contribute to this bulk two point function.
\begin{equation}
\label{phiphiembedding}
\begin{aligned}
	\expval{\phi(x,y)\phi(0,y')} 
	&= \frac{\kappa}{(P_1\cdot V P_2\cdot V)^{\Delta_1}}\left( u^{\Delta_1}+\chi\left(\frac{u}{1-4u}\right)^{\Delta_1}\right) \\
	&=\frac{1}{(P_1\cdot V P_2\cdot V)^{\Delta_1}}\left( \frac{C_{\expval{\phi \varphi_1}}^2}{C_{\expval{\varphi_1 \varphi_1}}}\hat{g}_{\frac{d}{2}-1}(u)+\frac{C_{\expval{\phi \varphi_2}}^2}{C_{\expval{\varphi_2 \varphi_2}}}\hat{g}_{\frac{d}{2}}(u)\right) \ . \\
\end{aligned}
\end{equation}
%
In this notation, the boundary conformal blocks $\hat g_{\Delta}(u)$ encode
 the contribution of a boundary operator of dimension $\Delta$ -- in this case either $\varphi_1$ or $\varphi_2$ -- and its descendants
to the two point function.\footnote{%
 We use the normalization $\hat g_{\frac{d}{2}-1} = \frac{1}{2} \left(u^{\Delta_1} + \left( \frac{u}{1+4u} \right)^{\Delta_1} \right)$ and
 $\hat g_{\frac{d}{2}} = \frac{1}{4 \Delta_1} \left(u^{\Delta_1} - \left( \frac{u}{1+4u} \right)^{\Delta_1} \right)$. 
} 
 In the case of either purely Neumann or purely Dirichlet boundary conditions, one of the blocks will drop out.
In our choice of normalization, the two-point function coefficients for general $\chi$ are 
\begin{equation}
\begin{aligned}
	C_{\expval{\phi \varphi_1}}&=C_{\expval{\varphi_1 \varphi_1}}=\kappa(1+\chi) \ ,  \\
	C_{\expval{\phi \varphi_2}}&=C_{\expval{\varphi_2 \varphi_2}}=2\Delta_1 \kappa(1-\chi) \ .  
\end{aligned}
\end{equation}
We pick a definite solution of $\chi^2=1$ to define these quantities at zero coupling. The goal of this subsection is to determine how the boundary coupling $g$ affects $\chi$ and changes $C_{\expval{\phi\varphi}}$  from its free-theory value. In general, starting from an initial value $\chi^2 = 1$, we demonstrate that
\begin{equation}
\label{chideltachi}
	\chi\delta\chi = -\frac{g^2}{2\Delta_1}\frac{8\pi^d}{\Gamma(\frac{d}{2}-1)^2}C_{\expval{\varphi \varphi}}C_{\expval{\sigma\sigma}}	\ .
\end{equation} 
Specific cases of this result were previously obtained in ref.\ \cite{Herzog:2017vj}, while this general form appears in ref.\ \cite{Behan:2020vy}. Our starting point is the perturbative expansion:
\begin{align}
	\expval{\phi(P_1)\phi(P_2)}_{g}=& \expval{\phi(P_1)\phi(P_2)} \\
	& +g^2 \int D^{p}X_1 X_2 \expval{\phi(P_1)\varphi(X_1)}\expval{\sigma(X_1)\sigma(X_2)}\expval{\varphi(X_2)\phi(P_2)}+ \mathcal{O}(g^4)  \ .\nonumber
\end{align}

The $O(g^2)$ term in the expansion is controlled by the embedding space integral: 
\begin{equation}
	\mathcal{I}=\int D^{p}X_1 D^{p}X_2\frac{(P_1\cdot V)^{\Delta_\varphi-\Delta_1}}{(-2P_1\bullet X_1)^{\Delta_\varphi}}\frac{1}{(-2X_1\bullet X_2)^{\Delta_\sigma}}\frac{(P_2\cdot V)^{\Delta_\varphi-\Delta_1}}{(-2P_2\bullet X_2)^{\Delta_\varphi}}	 \ .
\end{equation}
As will be discussed in sec.\ \ref{sec:shadowblocks}, because $\Delta_\sigma = p-\Delta_\varphi$, ${\cal I}$ is an example of the shadow integral used to compute conformal partial waves.  This connection explains why the correlator is manifestly conformally invariant:
these shadow integrals produce a sum of conformal blocks.  Moreover, the result will be shadow-symmetric, i.e.\ invariant under the replacement of 
$\Delta_\varphi$ with $p - \Delta_\varphi$, up to an overall normalization factor. In our context, the shadow symmetry occurs because the two operators entering the boundary OPE of $\phi$ satisfy the shadow relation $\Delta_{\varphi_1}+\Delta_{\varphi_2}=p$. 

We can do this integral through repeated use of Schwinger parametrization and the elementary formula of conformal integrals 
 (details in App.\ \ref{app:Int}). There remains a Schwinger integral which can be difficult to evaluate. By replacing the integrand by a power series, then integrating before resumming, we find that it is given by an hypergeometric function:
\begin{equation}
	\int_0^\infty  \frac{d\alpha}{\alpha} \frac{\alpha^{\Delta_\varphi}}{\left(\left(1+\alpha\right)^2+\frac{\alpha}{u}\right)^\frac{p}{2}} =\frac{\Gamma (\Delta_\varphi) \Gamma (p-\Delta_\varphi)}{\Gamma (p)}\, _2F_1\left(\Delta_\varphi,p-\Delta_\varphi;\frac{p+1}{2};-\frac{1}{4 u}\right) \ .
\end{equation}
For both values of $\Delta_\varphi$, this takes a simple rational form that decomposes nicely into the basis of boundary conformal blocks of the $\langle \phi \phi \rangle$ correlator. All in all, we obtain a $O(g^2)$ correction to the propagator given by:
\begin{equation}
\label{pertanswer}
	{\cal I} =\frac{g^2}{(P_1\cdot V P_2\cdot V)^{\Delta_1}} C_{\expval{\phi \varphi}}^2 C_{\expval{\sigma \sigma}} \left(\hat g_{\frac{d}{2}-1}(u)-2\Delta \hat g_{\frac{d}{2}}(u)\right) \frac{4^{\Delta_1}\pi^{p}\Gamma(\Delta_\varphi-\frac{p}{2})\Gamma(\frac{p}{2})}{\Gamma(\Delta_\varphi)\Gamma(p)} \ .
\end{equation}
This modification can be encapsulated in the shift (\ref{chideltachi}) given at the beginning of this subsection.

\subsubsection{Computation of $\expval{\tau DD}$} 

To compute $\langle \tau DD \rangle$, we start with the fact that a nonzero $\langle \sigma \sigma \rangle$ two-point function 
implies a nonzero $\langle \tau \sigma \sigma \rangle$ three point function in the decoupled boundary theory via the Ward identity (\ref{TOOtoOO}).  
Next, the $g \sigma \varphi$ interaction leads to a nonzero three point function $\langle \tau \varphi \varphi \rangle$.  Then finally, because the displacement
operator $D$ can be expressed as an object quadratic in the boundary limit $\varphi$ of the free field $\phi$, 
we can compute $\langle \tau D D \rangle$ from $\langle \tau \varphi \varphi \rangle$
using Wick's Theorem.

The technically challenging
 part of the computation is the determination of $\langle \tau \varphi\varphi \rangle$ from $\langle \tau \sigma \sigma \rangle$, and so we start here.
A spinning version of the star-triangle relation, which we 
review in App.\ \ref{app:SpinStar}, makes the calculation straightforward.
We want to compute the following object:
\begin{equation*}
	 \int D^{p}X_1 D^{p}X_2 \expval{\tau(P_3,Z)\sigma(X_1)\sigma(X_2)}\expval{\varphi(X_1)\varphi(P_1)}\expval{\varphi(X_2)\varphi(P_2)}	\ .
\end{equation*}
Stripping off some two-point function coefficients, the central integral is
\begin{equation}
\begin{aligned}
	I &= \int \frac{D^{p}X_2}{(-2 P_3\cdot X_2)^{\frac{p}{2}+1}(-2 X_2 \cdot P_2)^{\Delta_\varphi}} \times 
\\
& \hspace{1in} \times \int D^{p}X_1 \frac{V_{P_3;X_1,X_2}^2}{(-2 P_3\cdot X_1)^{\frac{p}{2}+1}(-2 X_1\cdot X_2)^{\Delta_\sigma-\frac{p}{2}+1}(-2 X_1 \cdot P_1)^{\Delta_\varphi} } \ ,
\end{aligned}
\end{equation}
where $V_{1;2,3}$ was defined in (\ref{hatGOO}). The numerator of the inner integrand can  be factorized in the form $S^{AB}X_{1 A}X_{1 B}$, with 

\begin{align}
	S^{AB} = 4(Z\cdot X_2 P_3{}^{A}-P_3\cdot X_2 Z^{A})(Z\cdot X_2 P_3{}^{B}-P_3\cdot X_2 Z^{B}) \ .
\end{align}
Hence, we  can use the $l=2$ spinning star-triangle identity to evaluate it. Since $S^{AB}$ is orthogonal to  both $X_2$ and $P_3$,  only one  term survives in the sum:
\begin{equation}
\begin{aligned}
	&\int D^{p}X_1 \frac{V_{P_3; X_1,X_2}^2}{(-2 X_1 \cdot P_1)^{\Delta_\varphi}(-2 X_1\cdot X_2)^{\Delta_\sigma-\frac{p}{2}+1}(-2 X_1 \cdot P_3)^{\frac{p}{2}+1}} \\
	&= \frac{V_{P_3; P_1,X_2}^2}{(-2 P_3\cdot P_1)^{p-\Delta_\sigma+1}(-2P_3\cdot X_2)^{\frac{p}{2}-\Delta_\varphi}(-2P_1\cdot X_2)^1}\mu_2\left(\Delta_\varphi,\Delta_\sigma-\frac{p}{2}+1,\frac{p}{2}+1;2,0,0\right)
	\end{aligned}
\end{equation}
where the normalization factor $\mu_2$  (\ref{muell}) is defined in the appendix.

The remaining $X_2$ integral is structurally identical to the $X_1$ one we just performed. 
Only one term will survive.  The final result is
\begin{equation}
	\begin{aligned}
		I &= \frac{V_{P_3; P_1,P_2}^2}{(-2 P_3\cdot P_1)^{\frac{p}{2}+1}(-2 P_3\cdot P_2)^{\frac{p}{2}+1}(-2 P_1\cdot P_2)^{\Delta_\varphi-\frac{p}{2}+1}}\\
		&\times\mu_2\left(\Delta_\varphi,\Delta_\sigma-\frac{p}{2}+1,\frac{p}{2}+1;2,0,0\right)\mu_2(1,\Delta_\varphi,p-\Delta_\varphi+1;0,2,0) \ .
	\end{aligned}
\end{equation}
We recognize the standard three point function of a spin-2 primary of dimension $p=d-1$ with two identical scalar fields of dimension $\Delta_\varphi$.  The nontrivial result of our computation is the overall normalisation factor. One can appreciate how easy these manipulations were made using embedding space and conformal integrals; 
we found a  real space approach much more challenging.

Using this result, it is now a straightforward exercise in Wick contraction to obtain the result for $\langle \tau DD \rangle$. 
We find at leading order in $g$
\begin{equation}
\label{CtauDD}
	C_{\expval{\tau DD}}=-\frac{g^2}{4\Delta_1} \pi\frac{\Gamma (d+1)}{\kappa (2\pi)^d}C_{\expval{\varphi\varphi}}C_{\expval{\sigma \sigma}}+\mathcal{O}(g^4)  \ . 
\end{equation}

\subsection{Ward Identity Away From Free Theory}

We have now computed all the pieces necessary to verify our Ward identity \eqref{eq:ward}
to leading order away from free boundary conditions. What remains to check is 
\begin{align*}
	\underbrace{\frac{2}{d}S_{d-1}\int_{0}^{1}\left[\alpha(x^2)\left(d x^2-1 \right)+\beta(x^2)\left(x^2(2-d)-1 \right)
\right] \sqrt{1-x^2}^{d-3}dx}_{A}+& \\  \underbrace{\frac{d-2}{d-1}S_{d-1}C_{\expval{\tau DD}}}_{B}+\underbrace{d C_{\expval{DD}}}_{C}&=0 \ .
\end{align*}

From (\ref{cDDpredicted}), (\ref{CtauDD}), and (\ref{cDDactual}), we find that for general $\chi$ 
\begin{equation}
	\begin{aligned}
		A &= -d\frac{\Gamma \left(\frac{d}{2}\right)^2}{8\pi^d} \left(3 \chi ^2+1\right)  \ ,\\
		B &= - g^2 d \Delta_1 C_{\expval{\varphi\varphi}}C_{\expval{\sigma \sigma}} \ , \\
		C &= d\frac{\Gamma \left(\frac{d}{2}\right)^2}{4\pi^d} \left(\chi ^2+1\right) \ .
	\end{aligned}
\end{equation}
We now 
specialise the perturbative case, $\chi^2 \rightarrow 1+ 2\chi \delta \chi$, and employ (\ref{chideltachi}): 
\begin{equation*}
	A+C = -d\frac{\Gamma \left(\frac{d}{2}\right)^2}{4\pi^{d}}  \chi \delta  \chi = g^2d\Delta_1 C_{\expval{\varphi\varphi}}C_{\expval{\sigma\sigma}} \ .
\end{equation*}
Indeed, we see that $A+B+C = 0$ at leading order in the small $g$ expansion, as predicted
by the Ward identity (\ref{eq:ward}).  Before, we had verified it for some free theories, but now 
we have verified the case of an interacting theory as well.

\subsection{Generalised Free Theory Perturbation}

Our explicit computation showed how the boundary localised interactions made the boundary conditions shift away from $\chi^2=1$. It is a natural question then, whether such an interaction can produce a smooth interpolation from Dirichlet to Neumann. We here show how, by restricting $\sigma$ to be a generalized free field (GFF), we can resum all corrections to $\expval{\phi\phi}$ and obtain arbitrary $\chi$. This type of perturbation is akin to a large-$N$ expansion, and so our resummation result is similar to the one of \cite{Di-Pietro:2021wb}. We however perform the computation purely in position space, making use of the conformal integrals. We start by considering the full two point function:
\begin{equation}
\label{chaindiagrams}
\begin{aligned}
	\expval{\phi(P_1)\phi(P_2)}_g 
	&=\sum_{k=0}^{\infty} \frac{g^2}{(2k)!} \int \prod_{i=1}^{k}  D^{p}X_i \expval{\phi(P_1)\phi(P_2)\varphi(X_1) \ldots \varphi(X_{2k})}\expval{\sigma(X_1) \ldots \sigma(X_{2k})} \ . 
\end{aligned}
\end{equation} 
The odd terms in $k$ drop out because $\langle \varphi \rangle$ vanishes.  In fact,
because $\varphi$ and $\sigma$ are GFF, the summands reduce to chain-diagrams: 
\begin{equation*}
	\begin{aligned}
	\expval{\phi(P_1)\phi(P_2)}_g &=\sum_{k=0}^{\infty} g^{2k}\int  D^{p}X_1 \expval{\phi_{l}(P_1)\varphi(X_1)} \\
	&\int D^{p}X_2 \expval{\sigma(X_1)\sigma(X_2)}\int D^{p}X_3 \expval{\varphi(X_2)\varphi(X_3)}\ldots \\
	&\ldots\int D^{p}X_{2k}\expval{\sigma(X_{2k-1})\sigma(X_{2k})}\expval{\varphi(X_{2k})\phi(P_2)} \\
	&= \sum_{k=0}^{\infty} g^{2k} \mathcal{I}_{k} \ .
\end{aligned}
\end{equation*}
Using Schwinger parameters, we can compute the $X_1$ and $X_2$ integrals

\begin{equation*}
	\begin{aligned}
		&\int  D^{p}X_1 D^{p}X_2 \expval{\phi(P_1)\varphi(X_1)} \expval{\sigma(X_1)\sigma(X_2)}\expval{\varphi(X_2)\varphi(X_3)}\\
	&= C_{\expval{\sigma\sigma}}C_{\expval{\phi \varphi}}C_{\expval{\varphi\varphi}}\frac{\pi^{p}\Gamma(\frac{p}{2}-\Delta_\sigma)}{\Gamma(\Delta_\varphi)}\frac{\Gamma(\frac{p}{2}-\Delta_\varphi)}{\Gamma(\Delta_\sigma)}\frac{1}{(P_1\cdot V)^{\Delta_1-\Delta_\varphi}}\frac{1}{(-2P_1\cdot X_3)^{\Delta_\varphi}} \\
	&=\rho\expval{\phi(P_1)\varphi(X_3)}	
	\end{aligned}
\end{equation*}
where we have defined the 
shadow symmetric coefficient: 
\begin{equation*}
	\rho \equiv C_{\expval{\sigma\sigma}}C_{\expval{\varphi\varphi}} \pi^{p} \frac{\Gamma(\frac{p}{2}-\Delta_\varphi)}{\Gamma(\Delta_\varphi)}\frac{\Gamma(\frac{p}{2}-\Delta_\sigma)}{\Gamma(\Delta_\sigma)} \ .
\end{equation*}

This partial computation gives a recursion relation which is easily iterated:
\begin{equation*}
	\mathcal{I}_{k}=\rho\mathcal{I}_{k-1}=\rho^{k-1}\mathcal{I} \ .
\end{equation*}
The only piece left to compute is the initial integral, $\mathcal{I}$, which is exactly the result of our previous leading order computation (\ref{pertanswer}). The perturbative corrections are now easy to resum: 
\begin{equation}
\begin{aligned}
	\expval{\expval{\phi\phi}}&=\expval{\phi\phi}+ \sum_{k=1}^{\infty}g^{2k}\rho^{k-1}\mathcal{I} \\
	&= \expval{\phi\phi} + \frac{g^2}{1-g^2 \rho }\mathcal{I} \\
	&= \frac{1}{(P_1\cdot V P_2\cdot V)^{\Delta}}\frac{C_{\expval{\phi \varphi}^2}}{C_{\expval{\varphi\varphi}}} \Biggl[ \hat g_{\Delta_\varphi}(u) + \\
	& + \frac{g^2}{1-g^2 \rho} C_{\expval{\varphi\varphi}}C_{\expval{\sigma\sigma}} \left( \hat g_{\frac{d}{2}-1}(u)-2\Delta_1 \hat g_{\frac{d}{2}}(u)\right) \frac{4^{\Delta_1}\pi^{p}\Gamma(\Delta_\varphi-\frac{p}{2})\Gamma(\frac{p}{2})}{\Gamma(\Delta_\varphi)\Gamma(p)}
	\Biggr]
\end{aligned}
\end{equation}

This result is an intuitive deformation of our leading order computation. For convenience, we can use a rescaled interaction coefficient, replacing
$g^2$ with $\lambda = -g^2 \rho$. 
In this notation, starting from Dirichlet (D) Boundary conditions we obtain a resummed propagator: 
\begin{equation}
\begin{aligned}
	\expval{\phi\phi}_{\lambda,D}=\frac{\kappa}{(P_1\cdot V P_2\cdot V)^{\Delta}}&\left(4\Delta \frac{1}{1+\lambda}  \hat g_{\frac{d}{2}}(u)+2\frac{\lambda}{1+\lambda} \hat g_{\frac{d}{2}-1}(u) \right) \ .
	\end{aligned}
\end{equation}
The result for Neumann (N) Boundary condition is obtained under the map $\lambda \rightarrow \frac{1}{\lambda}$ \ .
\begin{equation}
\begin{aligned}
	\expval{\phi\phi}_{\lambda,N}=\frac{\kappa}{(P_1\cdot V P_2\cdot V)^{\Delta}}&\left(2 \frac{1}{1+\lambda} \hat g_{\frac{d}{2}-1}(u) + 4\Delta \frac{\lambda}{1+\lambda}   \hat g_{d/2} (u) \right) \ .
	\end{aligned}
\end{equation}
Comparing with (\ref{phiphiembedding}), we can read off an effective reflection coefficient
\[\label{eq:chiN}
\chi = \frac{1-\lambda}{1+\lambda} \ .
\]

One can wonder how this result plays out with the Ward identity. Since $\sigma$ is a GFF, its entire OPE is known. It is made up of primaries of all even spin and dimension $2\Delta_\sigma+n$. For the case under consideration, there is no spin-2 primary of dimension $d-1$ in it. Hence no $\tau$ is present. Our Ward identity then forces $\chi^2=1$. This is the usual statement that Dirichlet and Neumann are the only two free boundary conditions. However, we can still allow arbitrary $\chi$ by adding a contribution to the bulk stress tensor. If we consider a pair of scalar fields obeying Dirichlet $\phi_D$ and Neumann $\phi_N$ boundary conditions, we can consider a mixing of their boundary condition, through a boundary interaction of precisely the form considered here $g \phi_N \partial_n \phi_D$, where we can alternately consider the boundary field
$\sigma$ to be either $\phi_N$ or $\partial_n \phi_D$, depending on our point of view.  From a classical field theory perspective, the variational principle will 
then tell us how to relate $g$ to the corresponding change in boundary conditions on $\phi_N$ and $\partial_n \phi_D$ induced by the boundary interaction. Our computation shows that this effect is still true for the correlators. This model is now consistent with the Ward identity, by noting that the stress tensor has two pieces, $T$ that we  considered, and $T'$ from the second bulk field. A quick computation, completely analogous to the  one we just performed, gives the perturbatively induced coupling between the two bulk fields: 

\begin{equation}
	\begin{aligned}
		\expval{\phi_N(P_1)\phi_{D}(P_2)}&=\frac{C_{\expval{\phi_N \phi_D}}}{(-2P_1\cdot P_2+4 P_1\cdot V P_2 \cdot V)^{\Delta_1}} \ , \\
		C_{\expval{\phi_N \phi_D}}&= \frac{\sqrt{\lambda}}{1+\lambda}2\kappa \ .
	\end{aligned}
\end{equation}
Hence, their two point function is a pure reflective term. From this result, we can compute the contributions from $\expval{T' DD}$, and plug it into our integral constraint. One then finds that the Ward identity is satisfied provided \eqref{eq:chiN} holds.

Of course $\sigma$ does not have to be a GFF.  In the models considered in ref.\ \cite{Herzog:2017vj, Di-Pietro:2021wb}, it was a composite field, 
for example a boundary fermion or scalar bilinear.  In this case, through a Hubbard-Stratonovich transformation, the composite field can be
traded for a GFF at leading order in a large $N$ expansion.  However, at subleading order in $1/N$, one will find other diagrams besides
the ones in (\ref{chaindiagrams}) contributing to $\langle \phi \phi \rangle$.  For example, treating a fermion bilinear $\bar \Psi \Psi = \sigma$ as a generalized
free field, one is perforce neglecting loops that involve a fermion and a scalar.

This doubled $\phi_D \phi_N$  scalar point of view makes manifest a duality.  We see that the system $(\phi_D, \phi_N, \lambda)$ is equivalent to the
system $(\phi_N, \phi_D, \lambda^{-1})$.
This duality was discussed in an equivalent AdS/CFT context \cite{Witten:2001ua} a number of years ago.  Our free scalar system can be transformed
to AdS space by a Weyl rescaling of the metric, after which the two massless scalars $\phi_D$ and $\phi_N$ generate the same $R \phi^2$ type mass term
but continue to have different boundary conditions -- the so-called ``alternate quantizations'' discussed in the AdS/CFT literature.  
As the argument in ref.\ \cite{Witten:2001ua}  works for any negative mass-squared scalar in $AdS$, one should be able to generalize our argument
above to the Weyl equivalent case of a free scalar with the ``conformal mass'' discussed in ref.\ \cite{Herzog:2019bom}.

Quite recently, this duality was also discussed in the large-$N$ limit \cite{Di-Pietro:2021wb} (in 4d) in the cases where 
the $\sigma$ field was a composite boundary field, either a boundary fermion bilinear $\bar \Psi \Psi$ or a term quadratic in a boundary scalar $\Phi$.
There is then a duality between a free Dirichlet scalar $\phi_D$ in the bulk  coupled to $N$ free scalars $\Phi$ on the boundary on the one hand and 
a free Neumann scalar in the bulk $\phi_N$ coupled to $N$ critical scalars $\tilde \Phi$ on the boundary on the other.  The critical scalars are scalars
at the Wilson-Fisher fixed point of the boundary 3d theory. Similarly, one can trade the $N$ boundary scalars for $N$ boundary fermions. 
Subleading corrections in $1/N$ introduce a $\beta$ function for the coupling $\lambda$, and may spoil the duality, at least away from fixed points 
$g = 0$ and $g \to \infty$ where $\beta = 0$. 

A truncated version of the duality was posited earlier in ref.\ \cite{Klebanov:2002ja}, again in the AdS/CFT context.  
Translating to the current framework, the statement is that a Dirichlet scalar in the bulk is dual to 
$N$ critical scalars on the boundary.  One wonders the extent to which this boundary duality underlies much of the power of the AdS/CFT correspondence.


\section{Conformal Block Approach to the Ward Identity}\label{sec:block}

In this section we investigate the constraint on $C_{\expval{DDD}}$ arising from eq.\ \eqref{eq:ward} in a generic recombined theory, where the 
boundary operator $\tau_{ij}$ gets an anomalous dimension.
 To do so, we use the conformal block expansion to rewrite the constraint as a linear equation for the CFT data of the stress tensor. We first write down 
the  form of the expansion for a generic  bulk--boundary--boundary correlator in terms of eigenfunctions of the conformal Casimir \cite{Dolan:2001wg,Dolan:2004up,Dolan:2012wt}.
We then comment on the implication of our analysis on the anomaly coefficients.
The latter part of this section contains the technical details behind the derivation of the boundary block expansion.
 We showcase a weight-shifting operator which allows us to determine the scalar block entering $\expval{TDD}$ \cite{Costa:2011vf,Karateev:2018uk}. To find the tensorial block for spin-$2$ exchange, we  use the shadow formalism \cite{Ferrara:1972uy,Simmons-Duffin:2014wb}. 

\subsection{Casimir and Conformal Block}\label{sec:CPW}

Any correlation function in a CFT can be decomposed into a distinguished basis of functions. These are the conformal partial  waves (CPWs)\cite{Dolan:2004up,Dolan:2012wt}. They are characterised as eigenfunctions of the conformal Casimir operator. We will be concerned with the so-called boundary CPW, which encode the contribution to a correlation function coming from the BOPE decomposition in a channel consisting of a single primary field, specified by the labels $(L,\Delta)$ of its boundary spin and conformal weight. In our example, this expansion takes the form: 
\begin{equation}
\label{confsum}
\begin{aligned}
	\expval{T(P_1,Z)D(P_2)D(P_3)}&=\sum_{L,\Delta}\frac{C_{\expval{T \mathcal{O}}}C_{\expval{\mathcal{O}DD}}}{C_{\expval{\mathcal{O}\mathcal{O}}}}\mathcal{W}^{(2)}_{L,\Delta}(P_1,P_2,P_3;Z) \\
	&=\sum_{L,\Delta}\frac{C_{\expval{T \mathcal{O}}}C_{\expval{\mathcal{O}DD}}(\alpha_{L,\Delta}(u)(Y_2^2+Y_3^2)+2\beta_{L,\Delta}(u)Y_2 Y_3)}{{C_{\expval{\mathcal{O}\mathcal{O}}}}(-2P_1 \cdot P_2)^{\frac{d}{2}}(-2P_1 \cdot P_3)^{\frac{d}{2}}(-2P_2 \cdot P_3)^{\frac{d}{2}}} \ .
	\end{aligned}
\end{equation}
In the second line, we used the transitivity of the  conformal symmetry of the correlator onto the partial waves to re-express each of them in terms of a kinematical (tensor) part and a series of conformal blocks multiplying this tensor structure. Our goal will be to compute these functions and use their explicit form to write \eqref{eq:ward} in terms of the CFT data. 

From a Hilbert space perspective, the partial waves are derived by inserting a partition of unity which is organised by conformal family, 
i.e.\ a primary operator and all its descendants. 
Each conformal family gives a contribution to the correlation function which is encoded by one function, $\mathcal{W}^{(2)}_{L,\Delta}$. 
The partial wave naturally depends on the spin of the external bulk fields, which we indicated. In the situation we are interested in, we should think of inserting a complete set of boundary states in between the operator insertions of $T$ and $DD$, hence using the boundary OPE of $T$ \cite{Billo:2016vm}. Following Dolan's approach, this construction is equivalent to  the group-theoretic decomposition by eigenfunctions \cite{Dolan:2004up}.
Note from the discussion in section \ref{sec:bulkbrytwopoint}, we anticipate that the sum (\ref{confsum}) 
will only involve $L = 0$ and $2$.  Moreover, for $L = 0$ 
$\Delta$ is constrained to be $d$.  In fact, as discussed already above, the expectation is the $\tau_{ij}$ operator $(2, d-1)$ and vector operator $(1, d)$ will
recombine into a longer $(2, \Delta)$ multiplet with $\Delta > d-1$.

The Casimir operator is a quadratic differential operator built from the $SO(d+1,1)$ symmetry generators $L_{AB}$, which in the embedding space act on fields as 
\begin{equation}
	i [L_{AB}, \mathcal{O}_i(Z_i,P_i)]=\mathcal{L}^{(i)}_{AB}\mathcal{O}_i(Z_i,P_i) \ ,
\end{equation}
with differential operator $\mathcal{L}_{AB}$ given as in (\ref{eq:generator}). The reduced conformal symmetry $SO(d,1)$ of the boundary states translates into a restriction to unbroken generators $J_{AB}$ or 
equivalently ${\mathcal J}_{AB}$ where 
\begin{equation}
		\mathcal{J}_{AB} = \Pi^{\parallel A'}_{A}\Pi^{\parallel B'}_{B}\mathcal{L}_{A'B'} \ .
\end{equation}
and furthermore $\Pi^{\perp}_{AB} \equiv V_{A}V_{B}$ while $\Pi^{\parallel}_{AB} \equiv \eta_{AB}-V_A V_B$.
From these operators, we can define the quadratic Casimir of interest for us: %
\begin{equation}\label{eq:Casimir}
		\mathcal{C}= -\frac{1}{2}{\mathcal J}^{AB}{\mathcal J}_{AB} \ .
\end{equation}

Acting with $\mathcal{C}$ on the point $P_1$ where $T$ is inserted gives a nontrivial eigenvalue problem which singles out the contribution from a conformal family. 
Since this correlation function has two independent tensor structures, we obtain a set of two coupled second-order differential equations for $\alpha$ and $\beta$, which depend on the Casimir eigenvalue:
\begin{equation}
	C_{J,\Delta}^{(d-1)}=\Delta(\Delta-d+1)+J(J+d-3) \ .
\end{equation} 
The solutions to this system of differential equations are singled out by matching to the boundary OPE, which gives a normalisation as well as a leading power law behaviour. 
The discarded solutions are shadow blocks, corresponding to the exchange of an operator of dimension $\tilde{\Delta}=p-\Delta$.

The second order system is :
\begin{eqnarray}
4 v^2 (1-v) \alpha'' - 2v (-5 + 6v+ d(-1+2v)) \alpha' - 4 v \beta' + \nonumber \\
 + (  3d - d(4+d) v) \alpha -2 (2+d) \beta &=& C_{\Delta, j}^{(d-1)} \alpha \ ,
 \label{alphaeq}  \\
 4v^2 (1-v) \beta'' -4 v \alpha' - 2v(-5+6v +d(-1+2v))\beta' - 2 d \alpha + \nonumber \\
+ (4 + d(3-4v) - 4v -d^2 v) \beta &=& C_{\Delta, j}^{(d-1)} \beta \ .
 \label{betaeq}
\end{eqnarray}
This system can be solved directly to obtain the scalar block:
\begin{equation}
\label{scalarblock}
	\begin{aligned}
		\alpha_{0,d}(v)&=-\frac{\, _2F_1\left(\frac{d}{2},\frac{d}{2};\frac{d+3}{2};v\right)+(d-1-2(v-1)) \, _2F_1\left(\frac{d}{2},\frac{d+2}{2};\frac{d+3}{2};v\right)}{4 (d+1) (v-1)} \ ,  \\
		\beta_{0,d}(v)&=-\frac{\, _2F_1\left(\frac{d}{2},\frac{d}{2};\frac{d+3}{2};v\right)+(d-1) \, _2F_1\left(\frac{d}{2},\frac{d+2}{2};\frac{d+3}{2};v\right)}{4 (d+1) (v-1)} \ .
	\end{aligned}
\end{equation}
We will present a different derivation below, starting from the conformal block for exchange between scalar operators and a spin shifting operator. The solution for the $L=2$ blocks on the other hand takes the more involved form
\begin{equation}
\label{spinningblock}
	\begin{aligned}
		\alpha_{2,\Delta}(v)&= \frac{v^{\frac{\Delta-d-2}{2}}}{4}\left(\frac{d}{(d-2-\Delta)} \, _2F_1\left(\frac{\Delta }{2}+1,\frac{\Delta}{2};\Delta-\frac{d-1}{2}+1;v\right) \right.  \\
		&\left.+\frac{(\Delta +2) (d (v-1)-2 v+1)}{(d-1) (d-2-\Delta)} \, _2F_1\left(\frac{\Delta }{2}+2,\frac{\Delta}{2};\Delta-\frac{d-1}{2}+1;v\right) \right) \ , \\
		\beta_{2,\Delta}(v)&= \frac{v^{\frac{\Delta-d-2}{2}}}{4}\left(\frac{(2 v-1)}{ (v-1) (d-2-\Delta)}\, _2F_1\left(\frac{\Delta}{2}+1,\frac{\Delta}{2};\Delta-\frac{d-1}{2}+1;v\right) \right. \\
		&\left. -\frac{(d-\Delta -1) (d (v-1)+1)}{(d-1) (v-1) (d-\Delta -2)} \, _2F_1\left(\frac{\Delta }{2},\frac{\Delta }{2};\Delta-\frac{d-1}{2}+1;v\right)\right) \ .
	\end{aligned}
\end{equation}
We did not find this form for $\alpha_{2, \Delta}$ and $\beta_{2, \Delta}$ from a direct consideration of (\ref{alphaeq}) and (\ref{betaeq}).  Instead we used
a more constructive approach,
the shadow formalism.  We will discuss this more constructive approach below, but first let us apply the result to the problem under consideration.

\subsection{Relating  $C_{\expval{DDD}}$ and $C_{\expval{DD}}$}

By inserting the boundary conformal block decomposition of $\langle T DD \rangle$ into the Ward identity (\ref{eq:ward}), we find the following
algebraic constraint on $C_{\expval{DDD}}$ and $C_{\expval{DD}}$.
\begin{eqnarray}
\label{spinsum}
	\lefteqn{\frac{(d-2) \pi ^{\frac{d-1}{2}}}{\Gamma \left(\frac{d+1}{2}\right)}C_{\expval{\tau DD}} + d C_{\expval{DD}}=} \\
	&& \frac{\pi ^{\frac{d-1}{2}} \Gamma \left(\frac{d-1}{2}\right)}{\Gamma \left(\frac{d}{2}\right)^2}C_{\expval{DDD}}+ \sum_{\Delta_i >d-1}C_{\Delta_i} \frac{(d-2) \pi ^{\frac{d-1}{2}} (d (\Delta_i -1)-\Delta_i  (\Delta_i +1)+1) \Gamma \left(-\frac{d}{2}+\Delta_i +\frac{3}{2}\right)}{(d-1) (d-\Delta_i -2) (d-\Delta_i -1) \Gamma \left(\frac{\Delta_i }{2}+1\right)^2}
	\ .
	\nonumber
\end{eqnarray}
Note that by choosing an appropriate normalization of the spin-$2$ boundary primaries, we can set $C_{\Delta_i}=C_{\expval{t_i DD}}$ for
$\Delta_i > d-1$. The case $\Delta_i = d-1$ is more involved.  As noted in the discussion of bulk-boundary correlation functions in sec.\ \ref{sec:bulkbrytwopoint},
a conserved spin two current cannot couple to a bulk operator, $C_{\expval{\tau {\mathcal O}}}=0$.
Indeed, in the sum over spin-$2$ primaries, the term multiplying $C_{\Delta_i}$ diverges as $\Delta_i \rightarrow d-1$. 
We could however incorporate $\Delta_i = d-1$ in the sum while requiring a finite answer by setting $\lim_{\Delta_i \rightarrow d-1}C_{\Delta_i}= (\Delta_i-d+1)C_{\tau DD}$, which reproduces the contribution from the $\tau$ term. In practice, in a generic BCFT there is no conserved spin-$2$ boundary primary operator; 
hence one can ignore the $\tau$ term.

We are interested in the anomaly coefficients of recombined theories in $4D$. There is then no distinguished $\tau$, and we can specify $d=4$ and use (\ref{b1b2def}) 
to simplify the constraint as
\begin{equation}
\label{b1b2rel}
	\frac{30}{\pi^4}b_2=\frac{35}{4\pi^4}b_1+\sum_{\Delta_i>3}C_{\Delta_i}\frac{2 \pi ^{3/2} ((\Delta_i -3) \Delta_i +3) \Gamma \left(\Delta_i -\frac{1}{2}\right)}{3 (\Delta_i -3) (2-\Delta_i) \Gamma \left(\frac{\Delta_i }{2}+1\right)^2} \ .
\end{equation}
We see that both anomaly coefficients are generally independent of each other. Owing to recent work on the constraint of supersymmetry on defect anomaly coefficients \cite{Drukker:2021tw,Bianchi:2020tp}, one could hope that supersymmetry would  directly relate these coefficients. From our analysis, we see that to get a direct relation, we should ask that all contributions from the spin-$2$ sector to be related to the displacement operator data. Since this requires relating the CFT data of operators of spin $0$ and $2$, we would need at least $\mathcal{N}=4$ supersymmetry. However, it seems unlikely that this is enough, as that would be the same as saying that a single superconformal family appears in the boundary OPE of the stress-tensor supermultiplet. Generically, the CFT data of the spin-$2$ sector are new pieces of information that make the two anomaly coefficients distinct.  

In the absence of a direct relation between $b_1$ and $b_2$ in special theories, one might still hope that (\ref{b1b2rel}) could be used to produce an inequality.
An issue is that the $C_{\Delta_i}$ do not have a definite sign.  
Consider the free cases analyzed in this work, for which the $\alpha$ and $\beta$ functions that determined $\expval{TDD}$ 
were always simple polynomials of the cross ratio $v$, 
\begin{equation}
\begin{aligned}
\alpha(v) &= \frac{a_{-1}}{v} + a_0 - \frac{(d+4)(d-2) (d^2 a_0 + (d^2- 4) b_0)}{16(d-1)}v \ , \\
\beta(v) &= -\frac{a_{-1}}{v} + b_0 + \frac{d(d+2)(d^2 a_0 + (d^2-4)  b_0)}{16(d-1)} v\ , 
\end{aligned}
\end{equation}
for appropriate constants $a_{-1}$, $a_0$ and $b_0$.  Decomposing these functions into 
the conformal blocks, we find by inspection that
\begin{eqnarray}
C_{\expval{DDD}} &=& -\frac{4 a_{-1}}{d-1} + 2(a_0 + b_0) \ , \\
C_{d+2n} &=& a_{-1} \frac{2(-1)^n (d-1) \left(\frac{d}{2}+1 \right)_{n-1} \left(\frac{d}{2} \right)_{n+1}}{n! \left(\frac{d+1}{2}+n-1 \right)_{n+1}} \nonumber \\
&& +  \frac{(-1)^{n-1} n(1+d+2n)\left(\frac{d}{2}\right)_n \left(\frac{d+4}{2}\right)_{n-1}}{2d (4n+d-1) n! \left(\frac{d-1}{2}+n\right)_n } 
\biggl( a_0 d (4+d(n-1)(2n+d+3)) \nonumber \\
&& \hspace{0.5in}  + b_0 (d+2) (2-d(d+1)-2n+(d-1) dn + 2(d-2) n^2) \biggr) \ .
\end{eqnarray}
Indeed, even in this simple case, the coefficients are alternating in sign, making it difficult to turn (\ref{b1b2rel}) into an inequality.

\subsection{Blocks from the Shadow Formalism}
\label{sec:shadowblocks}

The shadow formalism takes advantage of the fact
that the conformal blocks are entirely fixed by symmetry; they are kinematical objects. From this insight, one can define a candidate CPW $\mathcal{W}_{L, \Delta}$ using conformal integrals to sew together lower point functions \cite{Ferrara:1972uy,Simmons-Duffin:2014wb}. In our situation, this takes the form:
\begin{equation}
\label{shadowintegral}
\begin{aligned}
		\mathcal{W}^{(2)}_{L,\Delta} &\sim \int D^{p}X \expval{T(P_1;Z)\hat{G}_{L,\Delta}^{a}(X)}\expval{\hat{G}_{L,\tilde{\Delta}}^{a}(X)D(P_2)D(P_3)}
		 \ ,
\end{aligned}
\end{equation}
where $\hat{G}_{L,\Delta}(X)$ is a boundary operator of dimension $\Delta$, and spin $L$, and the second insertion has dimension $\tilde{\Delta}=p-\Delta$. The notation follows that of \eqref{FJGhatL} and \eqref{hatGOO}. For a spinning primary, we need to fully contract indices between both operators, as we indicated by the repeated abstract index $a$. This ansatz can be motivated by noticing that the bulk-to-boundary correlation functions are natural eigenvectors of the Casimir operator acting on the bulk insertion. One can then get rid of the two new insertion points induced from the splitting by integrating over them, in a conformally invariant way. This expression can also be guessed from an Hilbert space perspective, in term of an insertion of the projector on the conformal family of a given boundary primary state. We omitted a normalization factor $\mathcal{N}$, which is fixed by requiring that the shadow of the shadow gives back the identity. In practice, we can fix the normalization by requiring the OPE matching. This representation has a shortcoming: it is invariant under $\Delta  \leftrightarrow \tilde{\Delta}$, and so gives a mixture of the block we want as well as the shadow block. Both contributions can be disentangled by performing a monodromy projection, as explained at length in ref.\ \cite{Simmons-Duffin:2014wb}. 

We compute the $(0,d)$ block using a weight-shifting operator and the explicit Casimir equation for a purely scalar three-point function. The $(2,\Delta)$ block we compute by directly evaluating the shadow integral (\ref{shadowintegral}).

\subsubsection*{Displacement Block through Recursion}

We exhibited a formal expression for the conformal blocks. Ref.\ \cite{Costa:2011vf} demonstrated how to use
an explicit expression for a lower spin conformal block 
to  obtain new results for higher spins through differential relations. In subsequent work \cite{Karateev:2018uk}, the procedure was formalized using the language of  weight-shifting operators, and was later applied to defect and boundary CFT in \cite{Lauria:2019wt}. 
We illustrate the method in our situation. Let us imagine that we have some differential operator $\mathcal{D}\left(P_1,Z,\pdv{}{P_1}\right)$, such that: 
\begin{equation}
	\expval{T(P_1,Z)\hat{\mathcal{O}}(P_2)}\equiv\mathcal{D}^2\left(P_1,Z,\pdv{}{P_1},\pdv{}{Z}\right)\expval{\mathcal{O}(P_1)\hat{\mathcal{O}}(P_2)} \ .
\end{equation}
Inserting this identity in the shadow integral (\ref{shadowintegral}), we obtain the relation:
\begin{equation}
	\begin{aligned}
		\mathcal{W}_{0,d}^{(2)}(P_1,P_2,P_3;Z)=\mathcal{D}^2\left(P_1,Z,\pdv{}{P_1},\pdv{}{Z}\right)\mathcal{W}_{0,d}^{(0)}(P_1,P_2,P_3) 
	\end{aligned}
\end{equation}
The CPW $\mathcal{W}_{0,d}^{(2)}$ is an eigenvector of the Casimir operator \eqref{eq:Casimir} acting at $P_1$,  
	\begin{align}
		(\mathcal{C}-C_{0,d}^{(d-1)})\mathcal{W}_{0,d}^{(0)}(P_1,P_2,P_3)&=0 \, .
	\end{align}

We then have two tasks, find $\mathcal{D}$ and find $\mathcal{W}_{0,d}^{(0)}$. To identify $\mathcal{D}$, let us look at a generic bulk--boundary two-point function
(\ref{FJGhatL}): 
\begin{equation}
	\begin{bmatrix}
		n, & m &; \  \Delta_1, & \Delta_2
	\end{bmatrix}=\frac{S_1{}^{n} S_{12}{}^{m} }{(P_1\cdot V)^{\Delta_1-\Delta_2}(-2 P_2\cdot P_1)^{\Delta_2}}
\end{equation}
where $n=J-L$, $m=L$, and $J\geq L$.  
The operator we seek should be manifestly transverse, and have weight zero in all points. 
Out of the available elements, this requirement singles out 
\begin{equation}
	\mathcal{D} = \left[(Z \cdot V)P_1 -(P_1\cdot V)Z \right]\cdot \left(\pdv{}{P}+\frac{Z \cdot V}{P_1\cdot V}\pdv{}{Z} \right) \ ,
\end{equation}
for which we can check 
\begin{equation}
	\mathcal{D}\begin{bmatrix}
		n, & m &; \  \Delta_1, & \Delta_2
	\end{bmatrix}= (\Delta_2+n+m) \begin{bmatrix}
		n+1, & m &; \  \Delta_1, & \Delta_2
	\end{bmatrix} \ .
\end{equation}
Solving the recursion relation yields
\begin{equation}
	\begin{bmatrix}
		n, & m &; \  \Delta_1, & \Delta_2
	\end{bmatrix} = \frac{\mathcal{D}^{n}}{(\Delta_2+m)_{n}}\begin{bmatrix}
		0, & m &; \  \Delta_1, & \Delta_2
	\end{bmatrix} \ .
\end{equation}

In the context of weight-shifting operators, ${\mathcal D}$ is a spin-shift operator, and there is a seed conformal block corresponding to each spin of the boundary exchanged operator.\footnote{This operator is proportional to the operator $\hat{\mathcal{D}}^{\bullet}_1$ considered in \cite{Lauria:2019wt}.} Indeed, we can use the scalar partial wave to compute the spin-2 partial wave, for boundary scalar exchange. 

We now compute $\mathcal{W}^{(0)}_{0,\Delta}$. We first parametrise the partial wave in terms of a partial block $h_{\Delta}(v)$,
\begin{align}
\mathcal{W}^{(0)}_{0,\Delta} \equiv \frac{h_{\Delta}(v)}{(-2P_1\cdot P_2)^{\frac{\Delta_1}{2}}(-2P_1\cdot P_3)^{\frac{\Delta_1}{2}}(-2P_2\cdot P_3)^{\Delta_2-\frac{\Delta_1}{2}}} \ ,
\end{align}
for which the Casimir equation takes the form,
\begin{equation}
\begin{aligned}
	h_{\Delta}(v) \left(-d\Delta_1+\Delta_1+\Delta_1^2 (-(v-1))\right)-\Delta(\Delta-d+1)h_\Delta(v)\\
	-2 v \left((d-2\Delta_1+2 (\Delta_1+1) v-3) h_\Delta'(v)2 (v-1) v h_\Delta''(v)\right)=0 \ .
	\end{aligned}
\end{equation}
Taking out the overall power-law given by the OPE matching $h(v)=v^{\frac{\Delta-\Delta_1}{2}}f(v)$, we find that $f(v)$ satisfies a hypergeometric equation. A unique solution is selected by matching with the boundary OPE : 
\begin{equation}
	h_{\Delta}(v)=u^{\frac{\Delta -\Delta_1}{2}} \, _2F_1\left(\frac{\Delta }{2},\frac{\Delta }{2};\Delta-\frac{p}{2}+1;u\right) \ .
\end{equation}
To find $\alpha_{0,d}(v)$ and $\beta_{0,d}(v)$ we now simply  have to plug this result for $h_{d}(v)$ in $\mathcal{W}^{(0)}_{0,\Delta}$, act on it with $\mathcal{D}^2$, and collect the terms multiplying the correct tensor structures.
We find (\ref{scalarblock}), which also satisfies the conservation constraint, 
as expected for the stress tensor. 

One can check that this result is consistent with the boundary OPE;
taking the boundary limit of this partial wave recovers a three point function of displacement operators with the right normalisation:
\begin{equation}
	\lim_{v\rightarrow 0}\mathcal{W}^{(2)}_{0,d}=\frac{(V\cdot Z)^2}{(-2P_1\cdot P_2)^{\frac{d}{2}}(-2P_1\cdot P_3)^{\frac{d}{2}}(-2P_2\cdot P_3)^{\frac{d}{2}}} \ .
\end{equation}
The structure $(V\cdot Z)^2$ comes with a choice of normalization of the boundary OPE coefficient, which fixes $\frac{C_{\expval{TD}}}{C_{\expval{DD}}}=\frac{d}{d-1}$\cite{McAvity:1993ul,McAvity:1995tm}.

\subsubsection*{Spinning Block from the Shadow Formalism}

The elegant method we used for the scalar block sadly does not work for the boundary spin $L=2$ blocks. 
The best approach we found to perform this computation is inefficient: we  evaluated the shadow integral (\ref{shadowintegral}) 
and then performed the monodromy projection to remove the shadow block. 
As the logic is similar to that presented in ref.\ \cite{Simmons-Duffin:2014wb} and the computations can largely be automated, 
we will be brief, including only some of the technical details in app.\ \ref{app:Int}.

The generic seed partial wave $\mathcal{W}^{(L)}_{L,\Delta}$ is computed from the sewing of the following bulk-to-boundary and boundary-boundary-boundary correlators :
\begin{equation}
	\begin{aligned}
		\expval{F_{L}(P_1,Z)\hat{G}_{L,\Delta}(X,W)} &= \frac{1}{(P_1\cdot V)^{\Delta_1-\Delta}(-2 P_1 \cdot X)^{\Delta}}S_{12}{}^{L} \ , \\
		\expval{\hat{G}_{L,p-\Delta}(X,W)\mathcal{\hat{O}}(P_2)\mathcal{\hat{O}}(P_3)} &= \frac{V_{X;2,3}{}^{L}}{(-2 X \cdot P_2)^{\frac{\tilde{\Delta}+L}{2}}(-2 X\cdot P_3)^{\frac{\tilde{\Delta}+L}{2}}(-2 P_2 \cdot P_3)^{d-\frac{\tilde{\Delta}+L}{2}}} \ ,
	\end{aligned}
\end{equation}
from which we define the vectors $R^{A}$ and $Q^{A}$:
\begin{equation}
\begin{aligned}
		S_{12} &= \frac{P_1 \bullet W X\cdot Z}{P_1\cdot X}-W\bullet Z = R \cdot W \ , \\
		V_{X;2,3} &= -2 \left[ (W \cdot P_2) (X \cdot P_3)-(W \cdot P_3) (X \cdot P_2) \right]= Q \cdot W \ .
\end{aligned}
\end{equation}
The total contraction of these two correlation functions involves a contraction with the projector $\pi_{(L)}^{a_1 \ldots a_{l}, b_1 \ldots b_L}$ over the totally symmetric traceless transverse tensors. Such a contraction is known to be given by a Gegenbauer polynomial \cite{Dolan:2001wg,Costa:2014rya}: 
\begin{equation}
\begin{aligned}
		\pi_{(L)}(R,Q)&=c_{L}(R^2 Q^2)^{\frac{L}{2}}C_{L}^{\alpha}\left(\frac{R \cdot Q}{\sqrt{R^2 Q^2}}\right)  \  , \\
		c_{L}&=\frac{L!}{2^{L}(\alpha)_{L}} \  , \; \; \;
		\alpha =\frac{d-1}{2}-1 \ . 
\end{aligned}
\end{equation}
Hence the general seed partial wave for this configuration has the formal expression: 
\begin{equation}
	\mathcal{W}^{(L)}_{(L, \Delta)} \sim \int D^{p}X \frac{c_{L}\sqrt{R^2 Q^2}^{L}C^{\alpha}_{L}\left(\frac{R\cdot Q}{\sqrt{R^2 Q^2}}\right)}{(P_1\cdot V)^{\Delta_1-\Delta}(-2 P_1 \bullet X)^{\Delta}(-2 X \bullet P_2)^{\frac{\tilde{\Delta}+L}{2}}(-2 X\bullet P_3)^{\frac{\tilde{\Delta}+L}{2}}(-2 P_2 \bullet P_3)^{d-\frac{\tilde{\Delta}+L}{2}}} \ .
\end{equation}

We ignored the overall normalization constant, which we can fix later on, as well as  the monodromy projection for now. To compute this thing, we can expand the numerator, and we will obtain a sum of non-trivial tensorial integrals of the type: 
\begin{equation}
	\mathcal{I}_{(n)} =\int D^{p}X \frac{(Z\bullet X)^{(n)}}{(-2 P_1 \bullet X)^{a}(-2 P_2 \bullet X)^{b}(-2  P_3\bullet X)^{c}} \ , 
\end{equation}
for some $n$, $a$, $b$, and $c$ and for some $Z^2=0$ vector. These integrals can be rewritten as a sum of simpler scalar integrals, following the method of the spinning star formula in app.\ \ref{app:SpinStar}:
\begin{equation}
\begin{aligned}
		\mathcal{I}_{(n)}=&\frac{\sqrt{\pi}^{p}\Gamma(\frac{p}{2}+n)}{\Gamma(a)\Gamma(b)\Gamma(c)}\times \sum_{s+q+r=n}\frac{n!}{s! q!  r!}(P_1\bullet Z)^{s} (P_2\bullet Z)^{q}(P_3\bullet Z)^{r}\mathcal{F}_{a+s,b+q,c+r}(P_i) 
		\end{aligned}
\end{equation}
where we defined the elementary integral 
\begin{equation}
\label{elementaryintegral}
	\mathcal{F}_{a,b,c}(P_i)= \int_{0}^{\infty}\frac{d\beta d\gamma}{\beta \gamma}\frac{\beta^{b}\gamma^{c}}{\Big(-(P_1+\beta P_2 + \gamma P_3) \bullet (P_1+\beta P_2 + \gamma P_3)\Big)^\frac{a+b+c}{2}} \ .
\end{equation}
Sadly, unlike in the $\expval{\tau DD}$ computation, in general all terms contribute to the sum.  

The elementary integral (\ref{elementaryintegral})  can be computed as follows. First, by expanding the denominator, we can perform the $\gamma$ integral, 
recognising a Schwinger parametrization. We can change variables to obtain an overall weight times an integral which only depends on the cross-ratio $v$: 
\begin{equation*}
\begin{aligned}
	\mathcal{F}_{a,b,c}(P_i)&=\left((-2P_1\cdot P_2)^{\frac{c-a-b}{2}}(-2P_1\cdot P_3)^{\frac{b-a-c}{2}}(-2P_2\cdot P_3)^{\frac{a-b-c}{2}}\right)  \\
	&\times \frac{\Gamma(b)\Gamma(\frac{a-b+c}{2})}{\Gamma(\frac{a+b+c}{2})}\int_{0}^{\infty}\frac{ d\gamma}{\gamma}\frac{\gamma^{b}}{(1+\gamma)^{a} (v+\gamma)^{h-a}} \ .
\end{aligned}
\end{equation*}

This last integral can be evaluated by splitting it in two pieces using a contour deformation, and subsequently throwing away one of the two pieces because of the monodromy projection.
Once the dust settles we find that we should assign
\begin{equation*}
\begin{aligned}
	\mathcal{F}_{a,b,c}(P_i) \rightarrow  &\left(P_{12}{}^{\frac{c-a-b}{2}}P_{13}{}^{\frac{b-a-c}{2}}P_{23}{}^{\frac{a-b-c}{2}}\right)\\ 
	&\times \frac{ \Gamma (\frac{a-b+c}{2}) \Gamma (\frac{a+b-c}{2})\Gamma(\frac{b+c-a}{2})}{\Gamma (\frac{a+b+c}{2})}\\ 
	&\times \, _2F_1\left(\frac{a-b+c}{2},\frac{a+b-c}{2};1-\frac{b+c-a}{2};v\right) \ ,
\end{aligned}
\end{equation*}
and we used the shorthand $P_{ij}=-2P_i\cdot P_j$. It is now a straightforward, albeit tedious, exercise to plug  these formulas together, and consider $L=2,\Delta_1=d$. We then divide by the kinematic weight factors, and regroup the  expressions multiplying each tensor structure to find the candidate blocks $\alpha(v)$ and  $\beta(v)$. These functions still need to be normalized. The boundary OPE leading term  implies 
\begin{equation*}
	\begin{aligned}
		\lim_{v\rightarrow 0}\alpha_{2,\Delta}(v)&=\frac{v^{\frac{\Delta-d-2}{2}}}{4}(+1+\mathcal{O}(v)) \ , \\
		\lim_{v\rightarrow 0}\beta_{2,\Delta}(v)&=\frac{v^{\frac{\Delta-d-2}{2}}}{4}(-1+\mathcal{O}(v)) \ .
	\end{aligned}
\end{equation*} 
Being able to reproduce this  behaviour for both functions using only one normalisation is a further consistency check on our computation. 
Finally, we find (\ref{spinningblock}) for the spinning conformal block for the stress tensor.

\section{Discussion}

In the context of boundary CFT, 
we derived a Ward identity (\ref{eq:ward}) that ``integrates out'' the stress tensor from the bulk-boundary three point function $\langle T_{\mu\nu}(x_1) \hat {\mathcal O}({\bf x}_2) \hat {\mathcal O}({\bf x}_3) \rangle$
to give the corresponding boundary two point function $\langle \hat {\mathcal O}({\bf x}_2) \hat {\mathcal O}({\bf x}_3) \rangle$.  
Through the boundary OPE, the integral can be replaced by a sum over spin two boundary operators exchanged between the stress tensor and the two scalars
$\hat {\mathcal O}({\bf x}_2)$ and $\hat {\mathcal O}({\bf x}_3)$.  Because of our interest in boundary contributions to the trace anomaly, we gave an explicit
expression for this sum in the special case (\ref{spinsum}) where the scalars were displacement operators $D({\bf x})$.  
Indeed, through the relation between the boundary anomaly coefficients and the displacement operator correlation functions (\ref{b1b2def}), 
this sum rule can be phrased as a constraint on the coefficients $b_1$ and $b_2$ (\ref{b1b2rel}).
 We were able to check our Ward identity (\ref{eq:ward}) in a number of special cases: free scalar, free fermion, 4d Maxwell field, and a scalar
 interacting with a generalized free field localized on the boundary.  
 
Because of the perturbative analysis we carried out, we left a spin two, dimension $d-1$ boundary operator $\tau$ in the statement of the Ward identity (\ref{eq:ward}).
As discussed in the text, in a generic boundary CFT such a boundary stress tensor should be absent.  It will not be conserved.  Its divergence will source
a normal-tangential component of the bulk stress tensor, and hence the dimension of $\tau$ should be larger than the unitarity bound $d-1$ for spin two operators.
In our perturbative analysis, however, this shift occurs at subleading order in the expansion, and so at leading order, we need to keep this $\tau$ in 
the Ward identity.
 

The original motivation of this work was to put constraints on the boundary anomaly coefficients $b_1$ and $b_2$.  
We found that their difference depends on a sum over 
the three point function coefficients $C_{\Delta_i} = C_{t_i DD}$ where $t_i$ are the boundary spin-2 primaries exchanged in a boundary OPE
of the three point function $\langle T_{\mu\nu}(x_1)D({\bf x}_2) D({\bf x}_3) \rangle$.  It would be interesting to see if the 
difference between $b_1$ and $b_2$ can be constrained further. One path would be to use crossing symmetry or reflection positivity for example. These methods have been successfully employed to bound the difference in the bulk
coefficients $a-c$ \cite{Hofman:2016awc,Hartman:2016lgu}. Although we argued that supersymmetry is not enough to fully fix these coefficients, perhaps as in the case of $a-c$, it can still provide stronger constraints. A second path would then be to explore more quantitatively the consequence of superconformal invariance on the boundary OPE, building on \cite{Drukker:2017dgn,Drukker:2017tm}.

Another interesting generalization of the work here would be to consider a free scalar in the bulk with a ``conformal mass'' \cite{Herzog:2019bom,Carmi:2018qzm}.  
One adds a $\mu^2 \phi^2 / z^2$ term to the Lagrangian density, where $z$ is the distance from the boundary and $\mu$ is the conformal mass.
Such a term is consistent with the conformal symmetry preserved by the boundary.   A more covariant way of introducing the same effect is to introduce an external field $J(z) = \mu^2 / z^2$ with this power law profile and Weyl weight two. Our Ward identity should  generalize to this case. Instead of finding boundary fields $\varphi_N$ and $\varphi_D$ with dimensions $(d-2)/2$ and $d/2$, a conformal mass will allow boundary scalars with arbitrary dimension $\Delta \geq (d-3)/2$ greater than the unitarity bound for the boundary theory. After a Weyl rescaling, the theory becomes that of a massive scalar in anti-de Sitter space, making connection to similar types of calculations done in the AdS/CFT literature \cite{Witten:2001ua}.

\section*{Acknowledgments}
We would like to thank N.~Drukker, M.~Probst, M.~Trepanier, and R.~Mouland for discussion. VS would like to thank J.~Penedones and M.~Meineri for teaching him CFT. 
 This work was supported in part by a Wolfson Fellowship from the Royal Society
  and by the U.K.\ Science \& Technology Facilities Council Grant ST/P000258/1.


\appendix

\section{Conformal Integrals}\label{app:Int}

\subsection{Definition}

We consider the following integral \cite{Simmons-Duffin:2014wb}:
\begin{equation*}
	\int D^{d}X f(X\cdot P, X\cdot Z) = \frac{2}{\operatorname{Vol}(GL(1,\mathbb{R}^{+}))}\int_{X^{0}>0} dX^0 dX^{d+1}d^{d}X^{i} \delta(X^2)f(X\cdot P, X\cdot Z)
	\ ,
\end{equation*}
with $Z$ a polarization vector with the property $Z^2=0=Z\cdot P$. This object is manifestly invariant provided that $f(X\cdot P, X\cdot Z)$ is homogeneous of weight $-d$ in the variable $X$:
\begin{equation*}
	f(\lambda P\cdot X,\lambda Z\cdot X)=\lambda^{-d}f(P\cdot X,Z\cdot X)
	\ .
\end{equation*}
In this configuration, we are free to gauge-fix away $X^{+}=1$, and use the Fadeev-Popov method to reduce this to an integral over real space. It is then straightforward to derive the following formula, which is the essential ingredient to compute such conformal integrals:
\begin{equation}\label{eq:fun}
	\int D^{d}X \frac{1}{(-2 Y\cdot X)^{d}}=\frac{\sqrt{\pi}^{d}\Gamma(\frac{d}{2})}{\Gamma(d)} \frac{1}{(-Y\cdot Y)^{\frac{d}{2}}} \ .
\end{equation}

In practice, we can always bring ourselves to such a situation by using Schwinger parametrization  to bring together multiple power-laws: 
\begin{equation}
	\frac{1}{A_1^{a_1}\ldots A_n^{a_n}}=\frac{\Gamma(\sum a_i)}{\prod\Gamma(a_i)}\int \frac{d\alpha_2\ldots d\alpha_n}{\alpha_2 \ldots \alpha_n}\frac{\alpha_2^{a_2}\ldots  \alpha_n^{a_n}}{(A_1+\alpha_2 A_2 +\ldots \alpha_n A_n)^{\sum a_i}} \ .
\end{equation}
Possible tensorial structures are taken into account by substituting the insertions of $X$ with derivatives acting on the power-law. This is illustrated during the computation of the spinning star formula.

\subsection{Star-Triangle Relation}

In conformal perturbation theory, one often encounters products of power-laws involving two insertions, which are to be integrated over. By conformal invariance, such an object  can be written as a usual conformal correlator. The main question is then to compute the net effect that this integration gives on the resulting  correlation  function. The most useful nontrivial example is the \textit{star} diagram. We use a diagrammatic notation: a power-law between two points  is a solid  line, with the weight on top: 
\begin{equation}
\includegraphics[width=0.2\linewidth,valign=c]{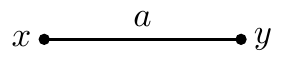} =\frac{1}{(x-y)^{2a}} \ .
\end{equation}

We will designate points which are integrated over using much bigger dots. The situation we are interested in is an insertion of a marginal deformation inside a three point function. This gives rise to a \textit{star diagram}:
\begin{equation}
\includegraphics[width=0.2\linewidth, valign=c]{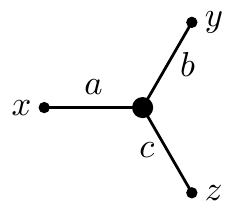} =\int d^{d}w\frac{1}{\abs{x-w}^{2a}\abs{y-w}^{2b}\abs{z-w}^{2c}}=I_{a,b,c} \ .
\end{equation}
These weights satisfy the identity $a+b+c=d$, which is crucial to our  next step. To compute this integral explicitly, we can use the usual methods of Feynman interals. However, we want to showcase a method which is both faster and more elegant, as well as adapted to the symmetry of the system. We uplift this integral to the embedding space: 
\begin{equation}
	\begin{aligned}
		I_{a,b,c}&=\int D^{d}X\frac{1}{(-2P_1 \cdot X)^{a}(-2P_2 \cdot X)^{b}(-2P_3 \cdot X)^{c}} \\
		&=\frac{\Gamma(d)}{\Gamma(a)\Gamma(b)\Gamma(c)}\int D^{d}X\int \frac{d\beta d\gamma}{\beta \gamma}\frac{\beta^{b}\gamma^{c}}{(-2(P_1 +\beta P_2+\gamma P_3)\cdot X)^{d}}\\
		&=\frac{\Gamma(d)}{\Gamma(a)\Gamma(b)\Gamma(c)}\frac{\sqrt{\pi}^{d}\Gamma(\frac{d}{2})}{\Gamma(d)}\int \frac{d\beta d\gamma}{\beta \gamma}\frac{\beta^{b}\gamma^{c}}{(-2(\beta P_1\cdot P_2 +\gamma P_1\cdot P_3+\beta \gamma P_2\cdot P_3))^{\frac{d}{2}}} \ .
	\end{aligned}
\end{equation}
Using Schwinger parametrization backward, we identify this as a product of power laws:
\begin{equation}
	\begin{aligned}
		I_{a,b,c}&=\frac{\sqrt{\pi}^d\Gamma(\frac{d}{2}-a)\Gamma(\frac{d}{2}-b)\Gamma(\frac{d}{2}-c)}{\Gamma(a)\Gamma(b)\Gamma(c)}\frac{1}{(-2P_1\cdot P_2)^{\frac{d}{2}-c}(-2P_1\cdot P_3)^{\frac{d}{2}-b}(-2P_2\cdot P_3)^{\frac{d}{2}-a}}\\
		&=\frac{\sqrt{\pi}^d\Gamma(\frac{d}{2}-a)\Gamma(\frac{d}{2}-b)\Gamma(\frac{d}{2}-c)}{\Gamma(a)\Gamma(b)\Gamma(c)}\frac{1}{\abs{x-y}^{d-2c}\abs{x-z}^{d-2b}\abs{y-z}^{d-2a}} \ .
	\end{aligned}
\end{equation}
We then define:
\begin{equation}
	\nu(a,b,c)=\frac{\sqrt{\pi}^d\Gamma(\frac{d}{2}-a)\Gamma(\frac{d}{2}-b)\Gamma(\frac{d}{2}-c)}{\Gamma(a)\Gamma(b)\Gamma(c)} \ ,
\end{equation}
from which we find the \textit{star-triangle formula}:
\begin{equation}
\includegraphics[width=0.2\linewidth, valign=c]{diagram2} =\nu(a,b,c)\includegraphics[width=0.3\linewidth, valign=c]{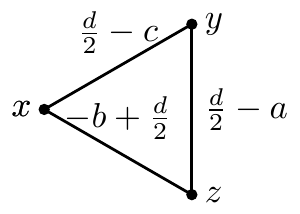}
\ . 
\end{equation}

It  may happen that one of the weights either equals $0$ or $\frac{d}{2}$. In  that case, we should regulate the computation by giving a small $\epsilon$ regulator to  the weight.  

\subsection{Spinning Star}\label{app:SpinStar}

Our computation necessitates a generalization of the previous formula. We consider a spinning diagram: 
\begin{equation}
\includegraphics[width=0.2\linewidth, valign=c]{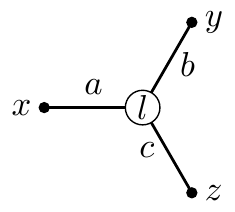} =\int D^{d}X\frac{(Z\cdot X)^{l}}{(-2P_1\cdot X)^{a}(-2P_2\cdot X)^{b}(-2P_3\cdot X)^{c}}=I^{l}_{a,b,c} \ .
\end{equation}
We decided to fully contract any free indices with a dummy and generic $Z^2=0$ vector. This way, we directly enforce from the start that the resulting object will be totally symmetric traceless at each step, and simplify the formula. We now have that $a+b+c=d+l$, as needed for conformal invariance. To compute this integral, we can proceed as follows : 
\begin{equation}
	\begin{aligned}
		I^{l}_{a,b,c}&=\frac{\Gamma(d+l)}{\Gamma(a)\Gamma(b)\Gamma(c)}\int D^{d}X\int \frac{d\beta d\gamma}{\beta \gamma}\frac{\beta^{b}\gamma^{c}(Z\cdot X)^{l}}{(-2(P_1 +\beta P_2+\gamma P_3)\cdot X)^{d+l}}\\
		&= \frac{\Gamma(d+l)}{\Gamma(a)\Gamma(b)\Gamma(c)}\frac{1}{2^l(d)_{l}}\int D^{d}X\int \frac{d\beta d\gamma}{\beta \gamma}\left(Z\cdot \pdv{}{Y}\right)^l\frac{\beta^{b}\gamma^{c}}{(-2Y\cdot X)^{d}}\\
		&= \frac{\Gamma(d+l)}{\Gamma(a)\Gamma(b)\Gamma(c)}\frac{\sqrt{\pi}^{d}\Gamma(\frac{d}{2})}{\Gamma(d)2^l(d)_{l}}\int \frac{d\beta d\gamma}{\beta \gamma}\left(Z\cdot \pdv{}{Y}\right)^l\frac{\beta^{b}\gamma^{c}}{(-Y\cdot Y)^{\frac{d}{2}}}\\
		&= \frac{\Gamma(d+l)}{\Gamma(a)\Gamma(b)\Gamma(c)}\frac{\sqrt{\pi}^{d}\Gamma(\frac{d}{2})\left(\frac{d}{2}\right)_l}{\Gamma(d)(d)_{l}}\int \frac{d\beta d\gamma}{\beta \gamma}\frac{(Z\cdot Y)^{l}\beta^{b}\gamma^{c}}{(-Y\cdot Y)^{\frac{d}{2}+l}}\\
		&= \frac{\Gamma(d+l)}{\Gamma(a)\Gamma(b)\Gamma(c)}\frac{\sqrt{\pi}^{d}\Gamma(\frac{d}{2})\left(\frac{d}{2}\right)_l}{\Gamma(d)(d)_{l}}\sum_{p+q+r=l}\frac{l!}{p!q!r!}(Z\cdot P_1)^{p}(Z\cdot P_2)^{q}(Z\cdot P_3)^{r}\int \frac{d\beta d\gamma}{\beta \gamma}\frac{\beta^{b+q}\gamma^{c+r}}{(-Y\cdot Y)^{\frac{d}{2}+l}}\\
		&= \frac{\sqrt{\pi}^d\Gamma(\frac{d}{2}+l)}{\Gamma(a)\Gamma(b)\Gamma(c)}\sum_{p+q+r=l}\frac{l!}{p!q!r!}(Z\cdot P_1)^{p}(Z\cdot P_2)^{q}(Z\cdot P_3)^{r}\mathcal{K}^{l}_{(a+p,b+q,c+r)}(P_i)  \ ,
	\end{aligned}
\end{equation}
with shorthand $Y=P_1+\beta P_2+\gamma P_3$. From this manipulation, we see clearly that this tensorial integral decomposes as a sum of polarizations times a given integral weight, which we now compute: 
\begin{equation}
	\begin{aligned}
		\mathcal{K}^{l}_{(a,b,c)}&=\int\frac{d\beta d\gamma}{\beta \gamma}\frac{\beta^{b}\gamma^{c}}{(-Y\cdot Y)^{\frac{a+b+c}{2}}} \\
		&=\int\frac{d\beta d\gamma}{\beta \gamma}\frac{\beta^{b}\gamma^{c}}{(-2(\beta P_1\cdot P_2 + \gamma P_1 \cdot P_3+\beta \gamma P_2\cdot P_3))^{\frac{a+b+c}{2}}} \\
		&=\frac{\Gamma(c)\Gamma(\frac{a+b-c}{2})}{\Gamma(\frac{a+b+c}{2})}\int\frac{d\beta}{\beta}\frac{\beta^{\frac{b+c-a}{2}}}{(-2P_1\cdot P_3-2 \beta P_2\cdot P_3)^{c}(-2 P_1\cdot P_2 )^{\frac{a+b-c}{2}}} \\
		&=\frac{\Gamma(\frac{a+b-c}{2})\Gamma(\frac{a-b+c}{2})\Gamma(\frac{b+c-a}{2})}{\Gamma(\frac{a+b+c}{2})} \times \frac{1}{(-2P_1\cdot P_2)^{\frac{a+b-c}{2}}(-2P_1\cdot P_3)^{\frac{a-b+c}{2}}(-2 \beta P_2\cdot P_3)^{\frac{b+c-a}{2}}}\\
	\end{aligned}
\end{equation}
We now define: 
\begin{equation}
\label{muell}
	\mu_l(a,b,c;p,q,r)=\frac{\sqrt{\pi}^{d}}{\Gamma(a)\Gamma(b)\Gamma(c)}\Gamma\left(\frac{d}{2}+l-c-r\right)\Gamma\left(\frac{d}{2}+l-b-q\right)\Gamma\left(\frac{d}{2}+l-a-p\right) \ .
\end{equation}
This gives us the \textit{spinning star triangle relation}:
\begin{equation}
\begin{aligned}
\includegraphics[width=0.2\linewidth, valign=c]{diagram4} &=\sum_{p+q+r=l}\frac{l!}{p!q!r!}\mu_{l}(a,b,c;p,q,r)\\
	&\times(Z\cdot P_1)^{p}(Z\cdot P_2)^{q}(Z\cdot P_3)^{r}\includegraphics[width=0.2\linewidth, valign=c]{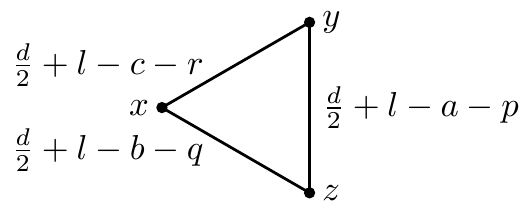}  \ .
\end{aligned}
\end{equation}
Of course, in this expression we can free-up indices using a Todorov operator to obtain a  \textit{bona fide} tensor, which will be totally symmetric traceless. 

\section{Free Fermion And 4d Maxwell Field}
\label{sec:free4d}

\subsubsection*{Free Majorana Fermion}

For a free Dirac fermion with Euclidean signature, we define the Clifford
algebra via $\{ \gamma_\mu, \gamma_\nu \} = 2 \delta_{\mu\nu}$.  The free Dirac equation
$\slashed{\partial} \psi = 0$ and conformal invariance forces the two point
function to take the form
\begin{equation}
\langle \psi(x) \psi^\dagger(x') \rangle = \kappa_f \left( \frac{  \gamma \cdot (x-x')}{|x-x'|^d} + \chi
\frac{ \gamma \cdot (\bar x - x')}{|\bar x - x'|^d} \right)
\end{equation}
where $\bar x = ({\bf x}, -y)$ and $\kappa_f^{-1} =S_d$.
In Euclidean signature, we take $\psi$ and $\psi^\dagger$ to be formally independent quantities with different boundary conditions
$\psi = \chi \psi$ while $\psi^\dagger = - \chi^\dagger \psi^{\dagger}$.  
For the two-point function to be compatible with the boundary conditions, we then require
\[
\chi \gamma_n =  \gamma_n \chi^\dagger \ , \;  \chi \gamma_i = -\gamma_i \chi^\dagger \ , \; \chi^2 = 1 \ ,
\]
and a convenient choice, which we will make, is to take $\chi = \gamma_n$.
The stress tensor takes the usual form
\begin{equation}
T_{\mu\nu} = \frac{1}{2} \left( (\partial_{(\mu} \psi^\dagger) \gamma_{\nu)} \psi - \psi^\dagger \gamma_{(\mu} \partial_{\nu)} \psi \right) \ .
\end{equation}
From this starting point, it is straightforward to work out the 
$\langle T_{\mu\nu}(x_1) D({\bf x}_2) D({\bf x}_3) \rangle$ and $\langle D({\bf x}) D({\bf x'}) \rangle$ correlation functions, being careful of the extra $-1$ from
the fermion loop.  We find that
\begin{equation}
\label{DDdirect}
\langle  D({\bf x}) D({\bf x'}) \rangle= \frac{2^{\lfloor \frac{d}{2} \rfloor} (d-1)  \kappa_f^2 }{|{\bf x} - {\bf x}'|^{2d}}
\end{equation}
along with the defining equations for $\langle T_{\mu\nu}(x_1) D({\bf x}_2) D({\bf x}_3) \rangle$:
\begin{equation}
\begin{aligned}
\alpha(v) &=- \frac{2^{\lfloor \frac{d}{2} \rfloor}  \kappa_f^3}{4v} d (d + (d^2-4) v)  \ , \\
\beta(v) &= \frac{2^{\lfloor \frac{d}{2} \rfloor} \kappa_f^3}{4v} d^2 (1 + d v) \ .
\end{aligned}
\end{equation}
Plugging these values for $\alpha$ and $\beta$ into the Ward identity (\ref{eq:ward}) yields the displacement two-point function coefficient
\begin{equation}
c_{\expval{DD}} = \frac{2^{\lfloor \frac{d}{2} \rfloor} (d-1) \Gamma \left(\frac{d}{2} \right)^2}{4 \pi^d} \ ,
\end{equation}
consistent with the direct computation (\ref{DDdirect}), assuming $c_{\expval{\tau DD}}=0$. 
\subsubsection*{Maxwell Field}

The Maxwell field is only conformal in 4d, and so we will restrict our analysis to that case.  
The stress tensor takes the usual form
\begin{equation}
T_{\mu\nu} =  F_{\mu \rho} {F_{\nu}}^\rho - \frac{1}{4} \delta_{\mu\nu} F_{\rho \sigma} F^{\rho \sigma} \ ,
\end{equation}
while for the two point function we take
\begin{equation}
\langle A_\mu(x) A_\nu(x') \rangle = \kappa \left( \frac{\delta_{\mu\nu}}{(x-x')^2} + \frac{\chi_{\mu\nu}}{(\bar x - x')^2} \right) \ .
\end{equation}
As usual $F_{\mu\nu} = \partial_\mu A_\nu - \partial_\nu A_\mu$.  The reflection coefficient $\chi_{\mu\nu}$ is equal to $\delta_{\mu\nu}$
up to a sign.  If the normal direction is chosen to be $\chi_{nn} = \pm 1$, then the tangential directions should be
set equal to $\chi_{ii} = \mp 1$(no summation implied).  These boundary conditions are sometimes called ``absolute'' and ``relative''.

With these choices we find
\begin{eqnarray}
\label{DDMaxwelldirect}
c_{\expval{DD}} &=& \frac{6}{\pi^4} 
\end{eqnarray}
along with the defining equations for $\langle T_{\mu\nu}(x_1) D({\bf x}_2) D({\bf x}_3) \rangle$:
\begin{eqnarray}
\alpha(v) &=& - \frac{8}{\pi^6} \frac{1}{v} \ , \; \; \;
\beta(v) = \frac{8}{\pi^6} \frac{1}{v} \ .
\end{eqnarray}
Plugging these equations into the Ward identity (\ref{eq:ward}), we recover the two point function coefficient (\ref{DDMaxwelldirect}), 
assuming $c_{\expval{\tau DD}}=0$. 

\bibliographystyle{jhep}
\bibliography{paper.bib}

\end{document}